\documentclass{aa}
\usepackage{amssymb,amsfonts,amsmath}
\usepackage[varg]{txfonts}
\usepackage{natbib}
\usepackage{graphicx}

\usepackage{bm}
\usepackage{color}
\usepackage{ulem}
\definecolor{mygray}{gray}{0.5}
\definecolor{magenta}{rgb}{0.858, 0.188, 0.478}

\usepackage{xspace}
\usepackage{upgreek}
\usepackage{enumitem}

\newcommand{\app}[1]{Appendix~\ref{app:#1}}
\newcommand{\App}[1]{Appendix~\ref{app:#1}}
\newcommand{\fg}[1]{Fig.~\ref{fig:#1}}
\newcommand{\Fg}[1]{Figure~\ref{fig:#1}}
\newcommand{\eq}[1]{Eq.~(\ref{eq:#1})}
\newcommand{\fgss}[2]{Figs.~\ref{fig:#1}--\ref{fig:#2}}

\newcommand{\Eq}[1]{Equation~(\ref{eq:#1})}

\newcommand{\se}[1]{Sect.~\ref{sec:#1}}

\newcommand{\Tb}[1]{Table~\ref{tab:#1}}

\newcommand{\ie}{i.e.,\xspace}
\newcommand{\eg}{e.g.,\xspace}

\newcommand{\dd}{\mathrm{d}}

\usepackage[T1]{fontenc}
\usepackage{CJKutf8}

\title{Catching drifting pebbles}
\subtitle{II. A stochastic equation of motion for pebbles}

\author{Chris W. Ormel, Beibei Liu }

\institute{Anton Pannekoek Institute (API), University of Amsterdam, Science Park 904,1090GE Amsterdam, The Netherlands\label{inst1} \\
\email{[c.w.ormel,b.liu]@uva.nl}
   }

\date{\today}

\abstract{Turbulence plays a key role in the transport of pebble-sized particles. It also affects the ability of pebbles to be accreted by protoplanets because it stirs pebbles out of the disk midplane. In addition, turbulence can suppress pebble accretion once the relative velocities become too high for the settling mechanism to be viable. Following Paper I, we aim to quantify these effects by calculating the pebble accretion efficiency $\varepsilon$ using three-body simulations. To model the effect of turbulence on the pebbles, we derive a stochastic equation of motion (SEOM) applicable to stratified disk configurations. In the strong coupling limit (ignoring particle inertia) the limiting form of this equation agrees with previous works. We conduct a parameter study and calculate $\varepsilon$ in 3D, varying pebble and gas turbulence properties and accounting for the planet inclination. We find that strong turbulence suppresses pebble accretion through turbulent diffusion, agreeing closely with previous works. Another reduction of $\varepsilon$ occurs when the turbulent rms motions are high and the settling mechanism fails. In terms of efficiency, the outer disk regions are more affected by turbulence than the inner regions. At the location of the H$_2$O iceline, planets around low-mass stars achieve much higher efficiencies.  Including the results from Paper I, we present a framework to obtain $\varepsilon$ under general circumstances.
}

\begin{document}
\begin{CJK*}{UTF8}{gbsn}
\maketitle

\section{Introduction}
It is widely believed that turbulence plays an important role in the evolution of protoplanetary disks.
For a long time the magneto-rotational instability (MRI) \citep{BalbusHawley1991} has been regarded as the leading candidate in driving the disk's angular momentum transport. More recently, disk wind models have regained traction \citep{BaiEtal2016,SuzukiEtal2016,Gressel2017}, where the turbulence in the midplane regions is limited to hydrodynamic instabilities such as the vertical shear instability \citep{NelsonEtal2013,StollKley2014}.
Turbulence, in addition, is important in shaping the outcome of the early coagulation process. Already at low mach numbers, turbulence dominates the relative velocity between particles \citep{VoelkEtal1980,OrmelCuzzi2007,PanPadoan2010}. It is also the only explanation for why we infer vertical structure (\eg\ flared versus settled geometry), since turbulence allows small particles to be lifted from the disk midplane regions. Indeed, with the current state-of-the-art models small (micron-size) particles are produced in the midplane through collisions between pebbles and boulder sized particles before they diffuse upwards \citep{BirnstielEtal2010i,BirnstielEtal2011,KrijtCiesla2016}.

Because of its subsonic nature, obtaining observational evidence of turbulence is hard. Disks like TW Hya, HD 163296, and DM Tau have been modeled by several groups \citep{HughesEtal2011,GuilloteauEtal2012,FlahertyEtal2015,FlahertyEtal2017} with turbulent Mach numbers inferred from the rather quiescent $\sim$0.01 to the more vigorous $\sim$$0.1$. However, it should be emphasized that constraining the turbulent rms velocity ($\sigma$) by these single-line profiles requires that the temperature profile be known to great precision. Generally, uncertainties affecting $\sigma$ are limited by the absolute flux calibration and spectral resolving power \citep{TeagueEtal2016}.  More indirect methods of obtaining $\sigma$ employ the appearance of the dust disk in ALMA imagery. Applied to HL tau this indicates that the pebbles are settled into the midplane, resulting in a vertical turbulent diffusivity parameter $\alpha_z\sim10^{-4}$ \citep{PinteEtal2016}.\footnote{The standard assumption is that $\alpha_z$ relates to the turbulent velocity as in $\sigma_z=\alpha_z^{1/2} c_s$ with $c_s$ the isothermal sound speed, but this identification assumes that the correlation time $t_\mathrm{corr}=\Omega^{-1}$. See discussion in \se{correlation}.} \citet{FlockEtal2017} conclude this is consistent with a magnetized disk models that feature a ``dead'' midplane.

Both classical planetesimal-driven models for planet formation \citep{Safronov1969,PollackEtal1996} as well as the more recent pebble accretion model \citep{OrmelKlahr2010,LambrechtsJohansen2012} are greatly affected by turbulence. Both models operate best under low-turbulence conditions.  The runaway growth phase for the classical, planetesimal-driven accretion paradigm can only operate once the planetesimals start out with close to zero velocity dispersions, but stochastic forcing by turbulence-triggered density fluctuations \citep{IdaEtal2008,NelsonGressel2010,GresselEtal2011,GresselEtal2012,OkuzumiOrmel2013} excites planetesimals to random velocities higher than their escape velocity. This implies that planetesimals have to be born large or that turbulence has to be weak \citep{OrmelOkuzumi2013,KobayashiEtal2016}. Similarly, the efficacy of pebble accretion to grow planets also depends on the turbulence. As pebbles will be stirred away from the midplane, it reduces the number of pebbles left to be accreted \citep{OrmelKlahr2010,GuillotEtal2014,MorbidelliEtal2015}. A second, less known, effect is that turbulent forcing may provide particles with an additional relative motion, which could also suppress accretion.

In this work, we consider simultaneously the effects of turbulent diffusivity and turbulent velocity.  We do this by deriving a stochastic equation of motion (SEOM) for pebble-sized particles. Simply put, the SEOM is an extension of the Newtonian equation of motion, but with an additional stochastic component. For planets, stochastic forces have been invoked as a means to cross mean motion resonances \citep{ReinPapaloizou2009,PaardekooperEtal2013}. As detailed by \citet{ReinPapaloizou2009} the stochastic motion is characterized by two parameters: the diffusivity $D_P$ and the correlation time $t_\mathrm{corr}$. The latter is crudely the time over which the stochastic force changes its direction. For pebbles, we adopt a similar model, where now the stochastic motions are driven by aerodynamical coupling to the turbulent gas.  However, in the few studies that have considered stochastic effects for pebble-sized particles, it is often assumed that turbulence does not feature a correlation time, \ie $t_\mathrm{corr}$ is assumed less than any other timescale in the problem \citep{Ciesla2010,ZsomEtal2011,KrijtCiesla2016}. This implies ``white noise'' behavior, \ie that the particle is displaced in a random direction at every time. This approximation is known as the strong coupling limit (SCA).

A key goal of this paper is to test how the SCA fares in the light of the more accurate SEOM. We find that the SCA is generally applicable, as long as both the particle stopping time $t_\mathrm{stop}$ and the turbulent correlation time are sufficiently small.  In addition, we will study the effect of a vertically varying turbulent gas diffusivity, $D_{zz}(z)$, to investigate when it is viable to stir a fraction of pebble size particles to the disk surface.

Our main thrust will be to apply the SEOM and the SCA methods to calculate pebble accretion efficiencies in three-dimensional (3D) settings. In \citet[][henceforth Paper I]{LiuOrmel2018} we have defined $\varepsilon$ as the probability that a pebble, drifting towards the star, will be accreted by a single planet(esimal)\footnote{In \citet{GuillotEtal2014} and \citet{LambrechtsJohansen2014} a similar quantity is defined.}. A very small value of $\varepsilon$ implies that a large number of pebbles are needed to grow the planet, while $\varepsilon$ close to unity implies that pebble accretion is a very efficient accretion process. In Paper I we used planar 3-body calculation (star, planet, pebble) to calculate $\varepsilon$ in two dimensions. We then investigated how this $\varepsilon_\mathrm{2D}$ changed as function of planet properties (mass and eccentricity), disk properties (position, radial drift velocity), and pebble properties (stopping time). In this work, we extend these calculation to the vertical dimension by including the planet's inclination and disk turbulence. With the $\varepsilon_\mathrm{3D}$ of this paper and the $\varepsilon_\mathrm{2D}$ of Paper I, we then obtain a general recipe for the pebble accretion efficiency ($\varepsilon$) of a single planet.

The plan of the paper is the following. In \se{model} we derive the SEOM. This section, as well as \app{Hottovy}, are rather technical and may be skipped by readers more interesting in the physical applications. In \se{test} we apply our newly developed SEOM to find vertical density distributions and show that our results are consistent with previous numerical and analytical studies.  We apply the SEOM and SCA to pebble accretion in \se{results}. We find $\varepsilon$ for a variety of settings (planet mass and inclination, particle and disk properties) and present a framework to obtain $\varepsilon$ under general circumstances (including the results found in Paper I). A comparison with previous studies is presented in \se{discussion}. We summarize our findings in \se{summary}.

\section{Model}
\label{sec:model}
\subsection{Advection-diffusion equation}
In this work we model turbulence motion of particles and gas by an advection-diffusion equation
\begin{equation}
    \frac{\partial\rho_P}{\partial t}
    + \nabla \cdot \rho_P \bm{v}
    - \nabla \cdot \rho_\mathrm{gas} \mathcal{D}_P \nabla \left( \frac{\rho_P}{\rho_\mathrm{gas}} \right)
    = 0
    \label{eq:advection-diffusion}
\end{equation}
where $\rho_P$ is the density of particles or gas species, $\bm{v}$ the systematic (drift) velocity, $\rho_\mathrm{gas}$ the gas density, and $\mathcal{D}_P$ the particle diffusivity tensor whose elements are denoted $D_{ij}$. Importantly, the diffusion term acts on the gradient of the concentration ($\rho_P/\rho_\mathrm{gas}$): it tends to erase concentration gradients and vanishes when the concentration is uniform.

In this work, we will restrict diffusion to operate only in the vertical ($z$) direction, considering only $D_\mathrm{P,zz}$. Furthermore, we adopt the vertically isothermal solution for the gas density
\begin{equation}
    \rho_\mathrm{gas} = \frac{\Sigma_\mathrm{gas}}{H_\mathrm{gas}\sqrt{2\pi}} \exp \left[ -\frac{1}{2} \left( \frac{z}{H_\mathrm{gas}} \right)^2 \right]
    \label{eq:rhogas}
\end{equation}
where $\Sigma_\mathrm{gas}$ is the gas surface density and $H_\mathrm{gas}$ the pressure scaleheight.  Under these conditions \eq{advection-diffusion} can be manipulated
\begin{equation}
    \frac{\partial\rho_P}{\partial t}
    + \frac{\partial}{\partial z} \left( v_z -\frac{D_\mathrm{P,zz} z}{H_\mathrm{gas}^2} \right)\rho_P
    = \frac{\partial}{\partial z} D_\mathrm{P,zz} \frac{\partial \rho_P}{\partial z}
    \label{eq:advection-diffusion2}
\end{equation}
\citep{Ciesla2010}\footnote{In \eq{advection-diffusion2} and other equations the differential operator $\partial/\partial z$ is understood to act on both terms to its right.}. For small particles (including pebbles) the vertical velocity $v_z$ equals the settling velocity, $v_z=-z\Omega^2 t_\mathrm{stop}$ with $t_\mathrm{stop}$ the stopping time and $\Omega$ the Keplerian orbital frequency.  The gas density no longer appears in \eq{advection-diffusion2}, but the diffusivity appears at two places. On the RHS the diffusive term is responsible for spreading the particle concentration, resulting in a broader distribution of $\rho_P(z)$. However, the additional advection term $-D_{\mathrm{P},zz} z/H_\mathrm{gas}^2$ -- a consequence of imposing \eq{rhogas} -- counteracts this, enforcing the particle layer to remain stratified with a finite dispersion at all times \citep{Ciesla2010}.

For a distribution of particles $\rho_P(z)dz/\Sigma$ gives the fraction of the particles within the interval $[z,z+dz]$. For a single particle, $P(z)=\rho(z)/\Sigma$ similarly denotes the probability of finding the particle within $[z,z+dz]$ where $P(z)$ is the probability density. We will use this identification below to obtain the correct, statistical properties of our single-particle (Lagrangian) stochastic model.

\subsection{Stochastic equation of motion (SEOM)}
The stochastic equation of motion is given by the following set of stochastic differential equations (SDEs):
\begin{subequations}
\label{eq:eom-all}
\begin{equation}
    \dd\bm{x} = \bm{v} \dd t
    \label{eq:eom-x}
\end{equation}
\begin{equation}
    \dd\bm{v} = \left( \bm{F}_g +\frac{-\bm{v} +\bm{v}_\mathrm{gas} +\sqrt{D_{zz}/t_\mathrm{corr}} \zeta_t\bm{e}_z  +\bm{v}_\mathrm{hs}}{t_\mathrm{stop}} \right) \dd t
    \label{eq:eom-v}
\end{equation}
\begin{equation}
    \dd\zeta_t = -\frac{\zeta_t}{t_\mathrm{corr}}\dd t +\sqrt{\frac{2}{t_\mathrm{corr}}} \dd W_t
    \label{eq:eom-zeta}
\end{equation}
\end{subequations}
where $\bm{x}$ is position and $\bm{v}$ the velocity of a particle. The particle is subject to gravitational forces $\bm{F}_g$ and gas drag forces. The latter have been expressed in terms of the stopping time, $\Delta\bm{v}/t_\mathrm{stop}$, where $\Delta\bm{v}$ is the relative gas-particle velocity. In \eq{eom-zeta} $W_t$ denotes a Wiener process (Brownian motion) and the corresponding differential is $\dd W_t\sim \sqrt{\dd t} \mathcal{N}(0,1)$ where $\mathcal{N}(0,1)$ is the normal distribution with zero mean and unity variance.

Apart from $\bm{v}$ three velocity terms appear in \eq{eom-v}:
\begin{enumerate}
    \item A laminar gas velocity $\bm{v}_\mathrm{gas}$. In our case, the gas velocity operates in the azimuthal direction
\begin{equation}
    \label{eq:vgas}
    \bm{v}_\mathrm{gas} = (1-\eta)v_K \bm{e}_\phi
\end{equation}
where $v_K=\sqrt{G(M_\star+M_p)/r}$ and $\eta$ represents the disk radial pressure gradient
\begin{equation}
    \label{eq:eta}
    \eta = -\frac{1}{2} \left( \frac{\partial \log P}{\partial \log r} \right)_\mathrm{midplane} \left(\frac{H_\mathrm{gas}}{r}\right)^2
\end{equation}
is assumed constant \citep{NakagawaEtal1986}.

    \item A turbulent velocity $\bm{v}_\mathrm{turb}$. In \eq{eom-v} this has been written in terms of an rms value ($\sigma_z = \sqrt{D_{zz}/t_\mathrm{corr}}$) and a non-dimensional stochastic variable $\zeta_t$. Here $t_\mathrm{corr}$ and $D_{zz}$ are, respectively, the correlation time and diffusivity of the turbulent gas. In \eq{eom-v} the turbulent forcing acts only in the vertical dimension.
    \item A correction term $\bm{v}_\mathrm{hs}$
        \begin{equation}
            \bm{v}_\mathrm{hs}
            =  -\frac{D_{zz} z}{H_\mathrm{gas}^2}\bm{e}_z +\frac{1}{2} D'_{zz}.
            \label{eq:vhs}
        \end{equation}
where $D'_\mathrm{zz}=\partial D_\mathrm{zz}/\partial z$. This is needed to enforce that \eq{eom-all} satisfies the hydrostatic balance condition, which assumption has entered the advection-diffusion \eq{advection-diffusion2}. It also accounts for spatial gradients in $D_{zz}$. We derive it below.
\end{enumerate}

Finally, \eq{eom-zeta} is a stochastic differential equation (SDE) about a quantity $\zeta_t$.  This can be thought of as the normalized strength of the turbulent velocity. Specifically, \eq{eom-zeta} describes an Ornstein–Uhlenbeck process \citep{UhlenbeckOrnstein1930} with zero mean ($\langle \zeta_t \rangle = 0$), unity variance ($\langle \zeta_t^2 \rangle = 1$) and correlation time $t_\mathrm{corr}$. On long timescales $\zeta_t$ will be normally distributed; events separated by $\Delta t \gg t_\mathrm{corr}$ are uncorrelated. However, times separated by $\Delta t \ll t_\mathrm{corr}$ will feature a similar value of $\zeta_t$ and hence a similar turbulent gas velocity.

\subsection{Turbulent correlation time}
\label{sec:correlation}
In \eq{eom-all} we are at liberty to choose $t_\mathrm{corr}$, which can be identified with the correlation time (or lifetime) of the turbulent eddies. A smaller $t_\mathrm{corr}$ (while keeping $D_{zz}$ fixed) implies a more vigorous turbulent forcing (larger $\sigma_z$), while a long $t_\mathrm{corr}$ implies that the turbulence is characterized by weaker but larger and longer-lived eddies. It is customary to adopt the \citet{ShakuraSunyaev1973} $\alpha$-parameterization for the turbulent viscosity
\begin{equation}
    \nu_T = \alpha H_\mathrm{gas}^2 \Omega.
    \label{eq:shakura-sunyaev}
\end{equation}
We will adopt a similar parameterization for the gas diffusivity, \ie $D_{zz} = \alpha_{z} H_\mathrm{gas}^2 \Omega$ where $\alpha_z$ reflects the diffusivity of the gas, not angular momentum transport. Using $D_{zz}=t_\mathrm{corr}^2 \sigma_z$ the turbulent rms velocity becomes
\begin{equation}
    \label{eq:sigma-z}
    \sigma_z
    = \sqrt{\frac{D_{zz}}{t_\mathrm{corr}}}
    = \frac{\alpha_z^{1/2} H_\mathrm{gas}\Omega} {\sqrt{t_\mathrm{corr} \Omega}}.
\end{equation}
Following \citet{DubrulleEtal1995}, \citet{CuzziEtal2001} and \citet{JohansenEtal2006i} we usually adopt $t_\mathrm{corr}=\Omega^{-1}$ and hence $\sigma_z=\alpha_z^{1/2} H_\mathrm{gas}\Omega$. In \se{massive} we also consider models where $t_\mathrm{corr}$ is longer.

The following qualifications will be adopted towards the turbulence strength:
\begin{itemize}
    \item laminar for $\alpha_z= 0$;
    \item weakly turbulent for $\alpha_z<10^{-4}$;
    \item moderately turbulent for $10^{-4}<\alpha_z<10^{-2}$;
    \item strongly turbulent for $\alpha_z>10^{-2}$.
\end{itemize}

\subsection{Strong coupling approximation (SCA)}
In the strong coupling approximation $t_\mathrm{stop}$ is assumed small. It can be shown that \eq{eom-all} then simplifies with the equation of motion being described by a single SDE
\begin{equation}
    \label{eq:xt}
    \dd\bm{x} = \left[ \bm{F}t_\mathrm{stop} +\bm{v}_\mathrm{gas} +\bm{v}_\mathrm{hs}
    +\frac{D'_\mathrm{zz}\mathbf{e}_z}{2(1+t_\mathrm{stop}/t_\mathrm{corr})}  \right] \dd t
    +\sqrt{2D_\mathrm{zz}} \mathbf{e}_z \dd W_t
\end{equation}
(see \App{Hottovy} for the derivation). In \eq{xt} we have allowed $D_\mathrm{zz}$ to depend on position, which would give rise to an additional advective term (the fourth term in the square brackets).  \Eq{xt} is analogous to \citet{Smoluchowski1916} equation for the stochastic motion of a massless particle subject to a fluctuating force.

We are now in a position to obtain the hydrostatic correction term $\bm{v}_\mathrm{hs}$. SDEs of the form
\begin{equation}
    \dd x = A(x) \dd t +B(x) \dd W_t
    \label{eq:xt3}
\end{equation}
can equivalently be cast in terms of an equation for the evolution of the probability density $P(x,t)$ -- \ie a Fokker-Planck equation
\begin{equation}
    \frac{\partial P(x,t)}{\partial t} +\frac{\partial}{\partial x} A(x)P
    = \frac{1}{2}\frac{\partial^2}{\partial x^2} B^2(x) P.
    \label{eq:FP-Ito-0}
\end{equation}
\citep{vanKampen1992}\footnote{\eq{FP-Ito-0} follows Ito's interpretation. See \se{Stratonovich} for the Stratonovich interpretation.}.  Applied to \eq{xt}, the Fokker-Planck equation for the probability density $P(z,t)$ reads
\begin{equation}
    \label{eq:FP-Ito}
    \frac{dP(z,t)}{dt}
    + \frac{\partial}{\partial z} \left( F_z t_\mathrm{stop}
    +\frac{1}{2}D_\mathrm{zz}' +v_\mathrm{hs} \right) P
    = \frac{\partial^2}{\partial z^2} D_\mathrm{zz} P
\end{equation}
where we consider the limit $t_\mathrm{stop} \ll t_\mathrm{corr}$.
The RHS of \eq{FP-Ito} can be expanded as $\frac{\partial}{\partial z} D_\mathrm{zz} \frac{\partial}{\partial z} P +\frac{\partial}{\partial z} D_\mathrm{zz}'P$.  Identifying the probability density $P(z)$ with the density $\rho_P$ of \eq{advection-diffusion2}, $F_zt_\mathrm{stop}$ with $v_z$, and using that $D_\mathrm{P,zz}=D_\mathrm{zz}$ for strongly coupled particles \citep{VoelkEtal1980}, we obtain the hydrostatic correction term, \eq{vhs}. With this correction term, the SCA for the particle position in 1D reads
\begin{equation}
    \dd z = \left[-z \Omega^2 t_\mathrm{stop} -\frac{D_\mathrm{zz} z}{H_\mathrm{gas}^2} +D_\mathrm{zz}'\right]\dd t +\sqrt{2D_\mathrm{zz}}\, \dd W_t.
    \label{eq:sca}
\end{equation}
as was already derived by \citet{Ciesla2010} and also used in \citet{ZsomEtal2011} and \citet{CharnozEtal2011}. Comparing the second and third terms on the RHS, we obtain that the turbulent gradient effect becomes important when $D_\mathrm{zz}$ changes on scales less than $\Delta z \sim D_\mathrm{zz}/D'_\mathrm{zz} = H_\mathrm{gas}^2/z$.

We reflect on our findings. \Eq{xt} with $\bm{v}_\mathrm{hs}$ equal to \eq{vhs} describes the stochastic motion of a particle experiencing drag and turbulent forces, with the turbulence characterized by a correlation time $t_\mathrm{corr}$ and a (possibly spatially dependent) gas diffusivity $D_\mathrm{zz}$. Under the assumption of small $t_\mathrm{stop}$ and small $t_\mathrm{corr}$ we obtain \eq{sca}, consistent with \citet{Ciesla2010}. However, these equations provide no model for the particle velocity; and they will fail when the SCA-conditions no longer materialize (long $t_\mathrm{stop}$ or long $t_\mathrm{corr}$) -- i.e., when the particle's inertia matters.  In these cases \eq{eom-all} provides a more general description of stochastic motion of pebble-sized particles.

\begin{figure}[t]
    \includegraphics[width=88mm]{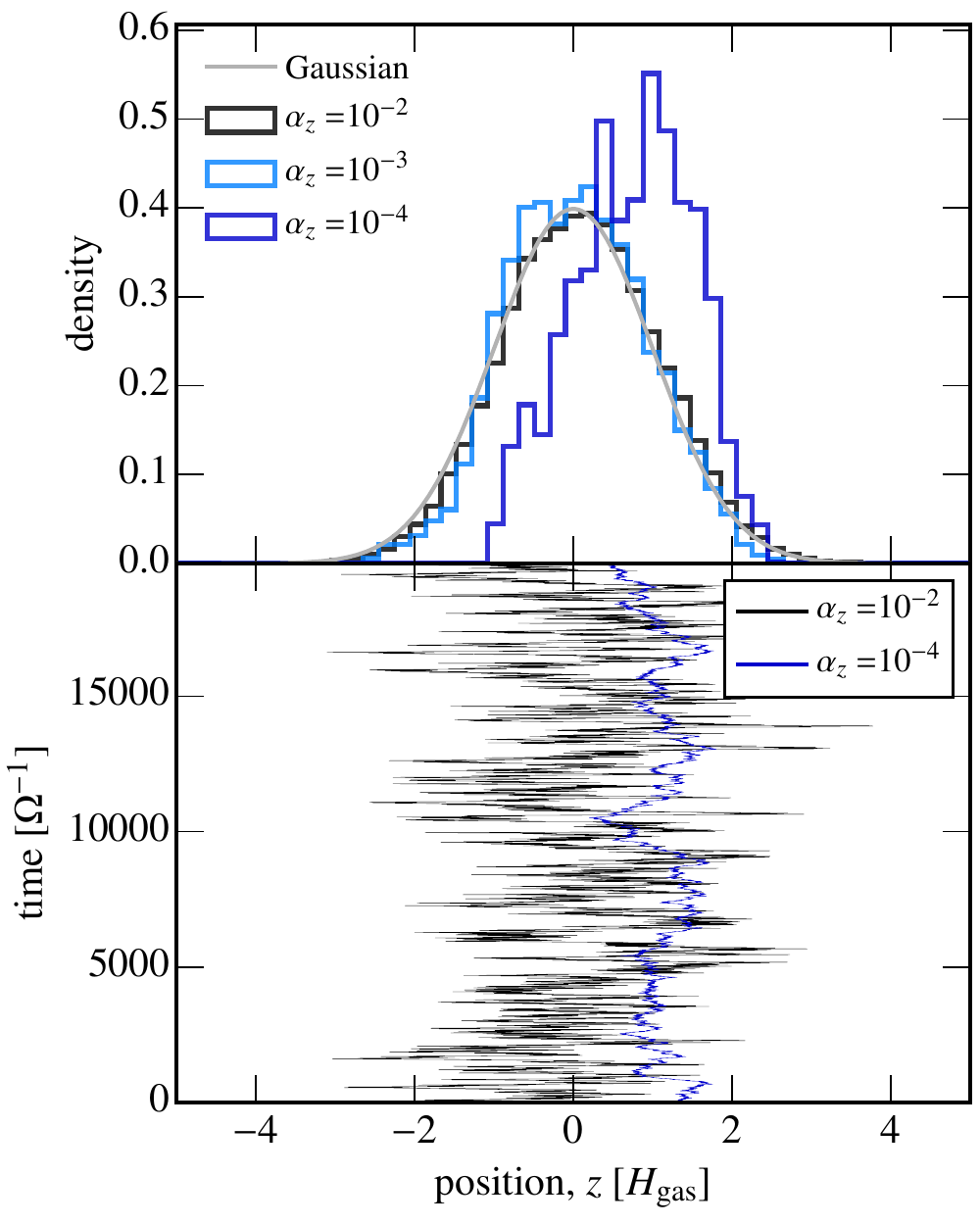}
    \caption{\textit{top}: Normalized distributions of the vertical position $z$ for the integration of the tracer case, using the strong coupling approximation (SCA) with $t_\mathrm{stop}=0$. The vertical height $z$ is recorded after every $1\,\Omega^{-1}$ for $t=10^5\,\Omega^{-1}$. Bars give the simulated distribution while the analytic -- normal -- distribution is shown by the black dashed line. \textit{bottom}: temporal evolution of the vertical position for $\alpha_z=10^{-2}$ (black) and $\alpha_z=10^{-2}$ (blue).
}
    \label{fig:tracers}
\end{figure}
\begin{figure*}[t]
    \includegraphics[width=0.33\textwidth]{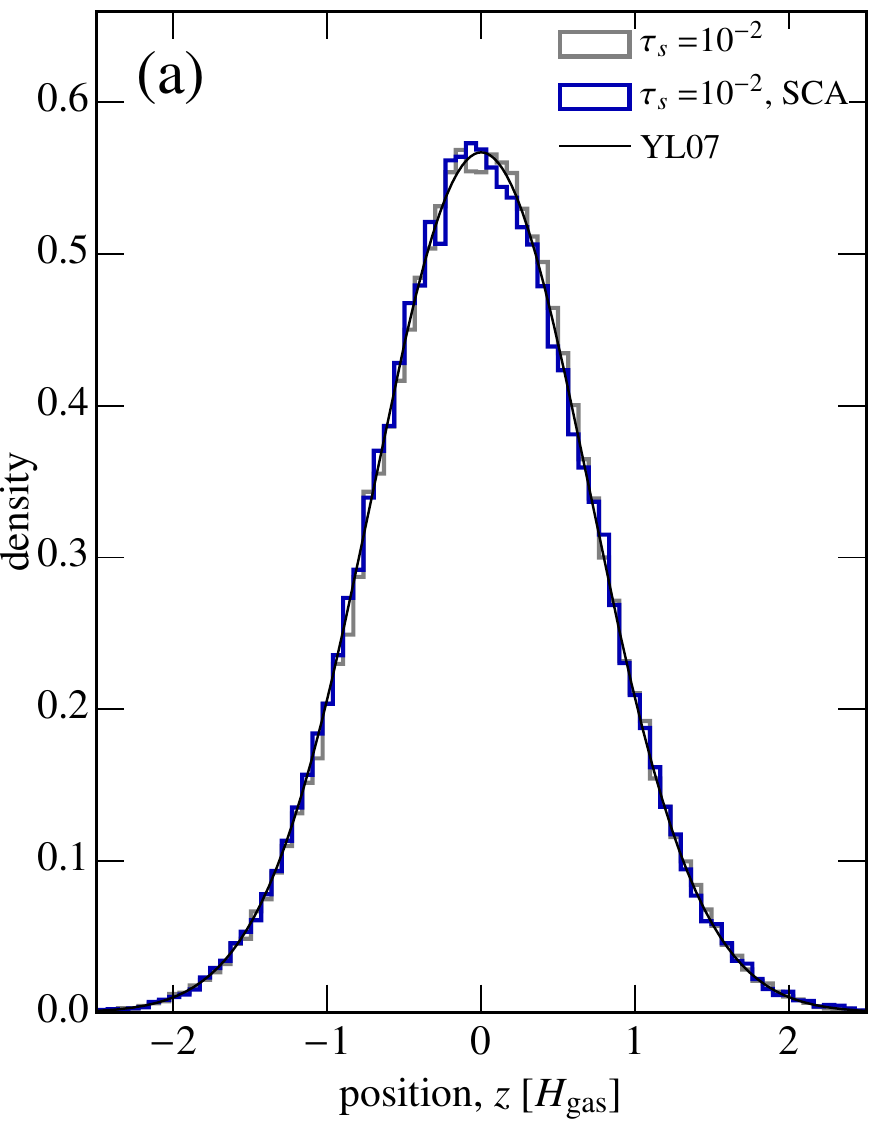}
    \includegraphics[width=0.33\textwidth]{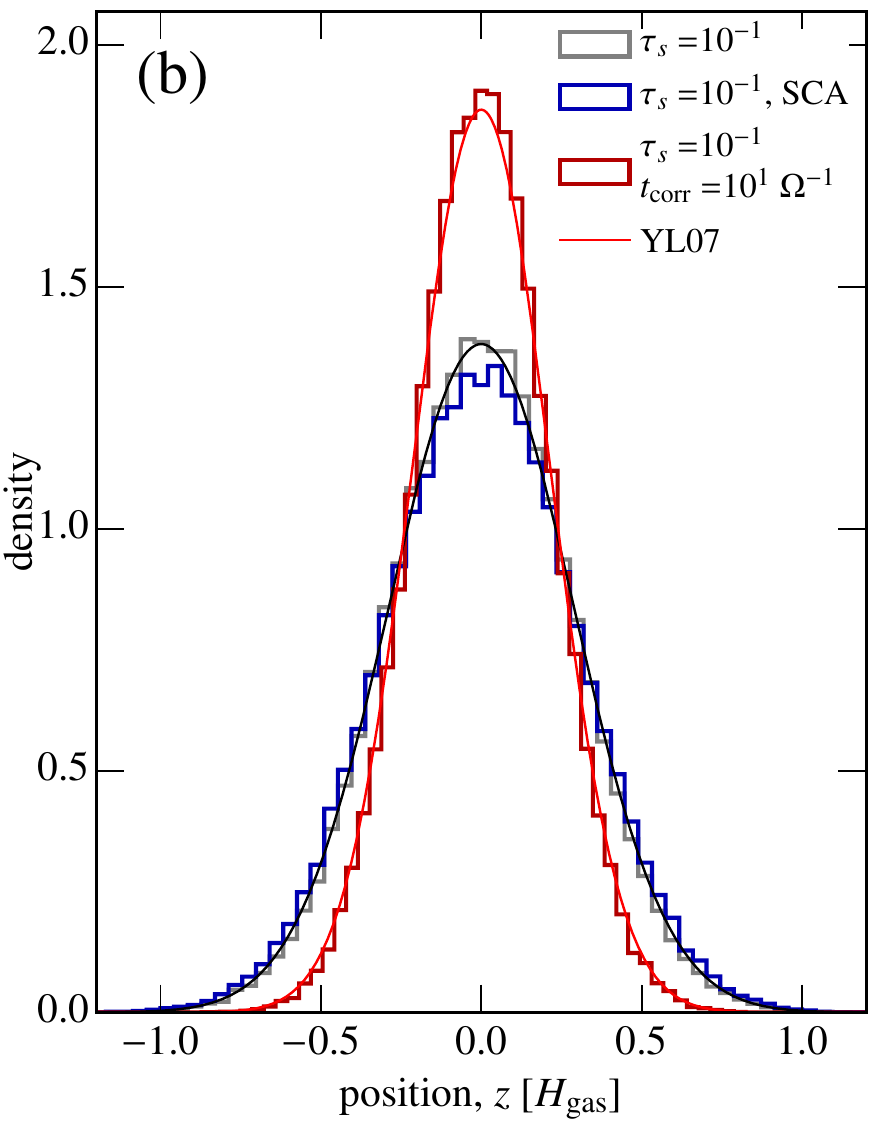}
    \includegraphics[width=0.33\textwidth]{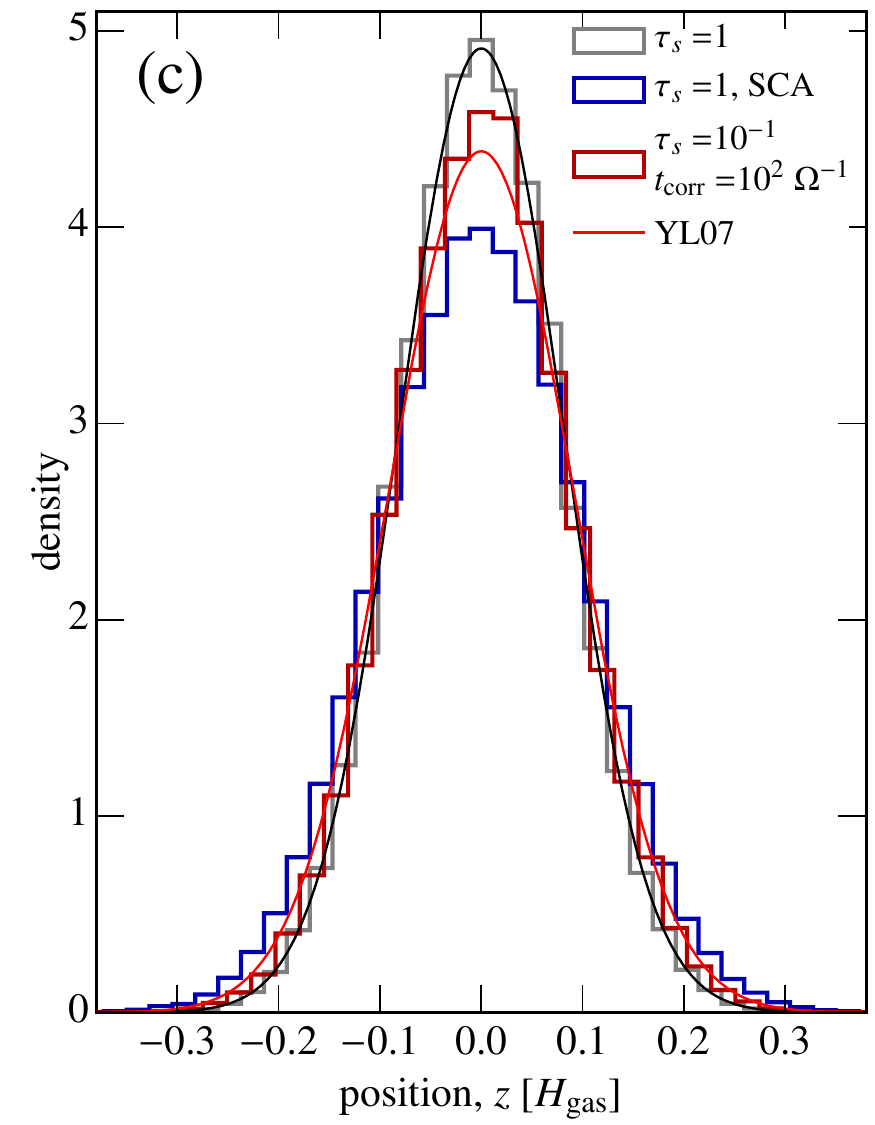}
    \caption{Vertical distribution for particles of different stopping times: $\tau_s=t_\mathrm{stop}\Omega=10^{-2}$ (\textit{left}), $10^{-1}$ (\textit{center}) and $1$ (\textit{right}). Histograms plot the numerically obtained distributions with the stochastic equation of motion (SEOM; gray and red) and strong coupling approximation (SCA; blue) methods. Long $t_\mathrm{corr}$ runs are shown with red histograms (the $t_\mathrm{corr}=10^2\,\Omega^{-1}$, $\tau_s=0.1$ run is displayed in panel c). Thin curves gives the normal distribution with the scaleheight of \eq{hp-YL07} \citep{YoudinLithwick2007}.  Note the different scaling among the panels.
    }
    \label{fig:particle-distr}
\end{figure*}
\begin{figure}[t]
    \includegraphics[width=0.46\textwidth]{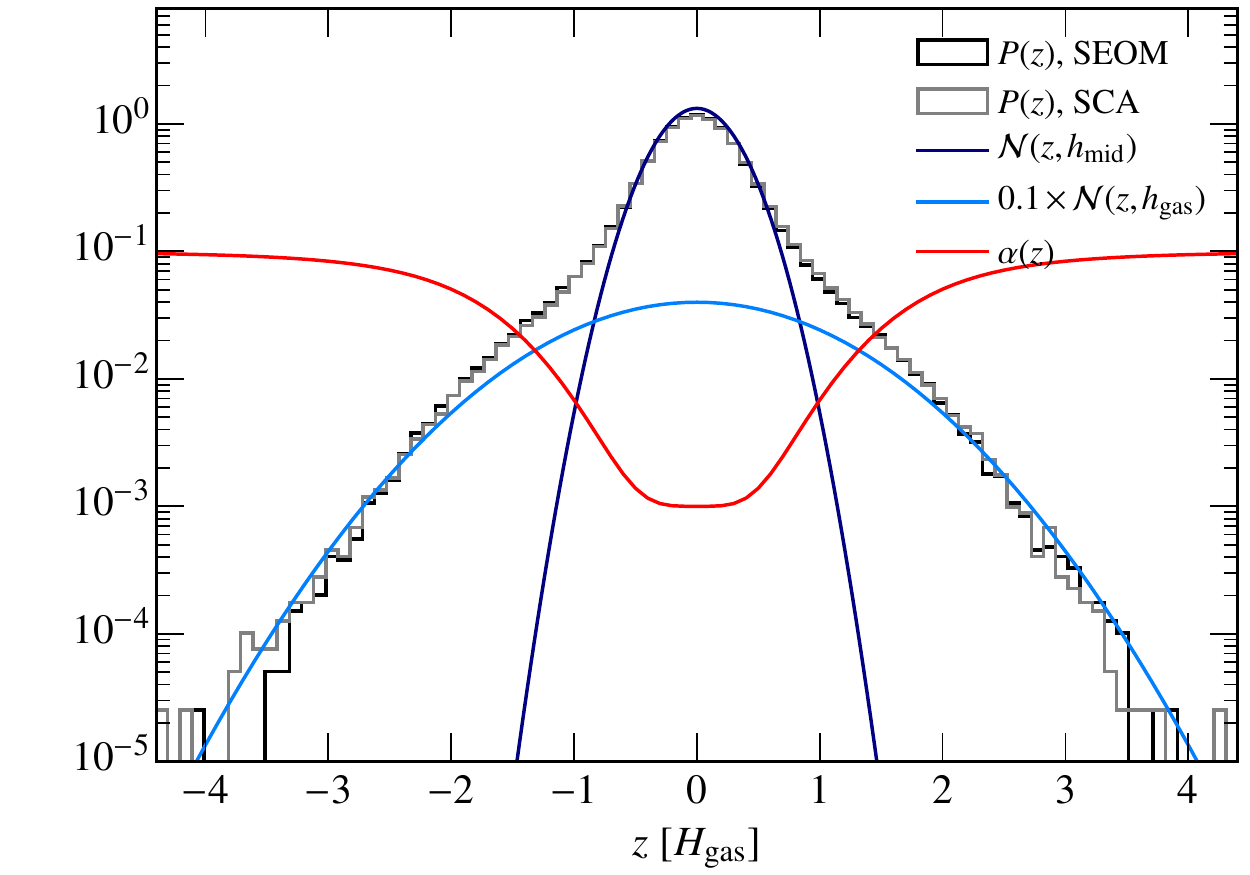}
    \caption{Density distribution $P(z)$ obtained from a vertically varying diffusivity. The diffusivity profile in terms of $\alpha_z$ is given by the red curve. The particle stopping time is $\tau_s=10^{-2}$ and is taken independent of $z$. The distributions obtained from integrating the stochastic equation of motions (\eq{eom-all}; black histogram) and the strong coupling approximation (\eq{xt}; gray histograms) are consistent. The dark blue curve gives the normal distribution, for the particle scaleheight evaluated in the midplane (\ie\ \eq{hp-YL07} with $\alpha=10^{-3}$). The light blue curves gives the gas distribution, scaled by a factor 0.1.}
    \label{fig:diffz}
\end{figure}

\section{Vertical diffusion}
\label{sec:test}
We test our algorithms -- the stochastic equation of motion (SEOM; \eq{eom-all}) and the strong coupling approximation (SCA; \eq{sca}) -- for tracer particles ($t_\mathrm{stop}=0$) in \se{tracer} and massive particles (\se{massive}) for a variety of stopping times and $\alpha_z$.

\subsection{Tracer particle}
\label{sec:tracer}
Tracer particles should have a vertical distribution identical to the gas, \eq{rhogas}. Because $t_\mathrm{stop}=0$ for tracer particles, the SEOM, as described in \eq{eom-all}, contains a singularity and is not applicable.  Therefore, we adopt the SCA method. In \eq{xt} we take $t_\mathrm{stop} = 0$, $\eta = 0$, $D_z = \alpha_z H_\mathrm{gas}^2 \Omega$. The choices for $H_\mathrm{gas}$ and $\Omega$ are arbitrary.

In \fg{tracers}a we show the distribution of the vertical position of a single particle for $\alpha_z=10^{-4}$, $10^{-3}$ and $10^{-2}$. These have been obtained by storing the vertical positions after every $1\,\Omega^{-1}$ for a total time of $t_\mathrm{max} = 10^5\,\Omega^{-1}$. In addition we plot the expected distribution according to \eq{rhogas}. The distributions are normalized such that they integrate to unity. Clearly, the distributions among the $\alpha_z$ differ, with $\alpha_z=10^{-2}$ best matching the expected normal distribution $\alpha_z=10^{-4}$ the worst. The origin of these differences is the different number of independent samples that are obtained among the $\alpha_z$. Since our sampling time is only $\Delta t = 1\,\Omega^{-1}$, much smaller than the diffusion time $t_\mathrm{diff} = 1/\alpha_z \Omega$, sequential samples (in time) will be strongly correlated; only $t_\mathrm{max}/t_\mathrm{diff} = 10^{5}\alpha_z$ samples will be independent. Hence, the higher $\alpha$, the better the correspondence to a Gaussian distribution.

\subsection{Massive particle}
\label{sec:massive}
Next we consider the vertical distribution of massive particles ($t_\mathrm{stop}>0$), obtained by the SEOM and the SCA methods. This means that in \eq{eom-all} we put $\bm{F}_g = -\Omega^2 z \bm{e}_z$. We further take $t_\mathrm{corr}=1\,\Omega^{-1}$ unless mentioned otherwise. Integrations ran for $10^6\,\Omega^{-1}$ and the sampling period was $\Delta t = 10\,\Omega^{-1}$. The results are shown in \fg{particle-distr}. We fix $\alpha_z=10^{-2}$ but vary $\tau_s=t_\mathrm{stop}\Omega=10^{-2}$ (left panel), $10^{-1}$ (center), and $10^0$ (right).

In all panels we compare the results for the SEOM (gray curves) with the SCA of \eq{xt} (blue).
The thin black curve corresponds to a normal distribution with pebble aspect ratio
\begin{equation}
    \label{eq:hp-YL07}
    h_P = \sqrt{\frac{\alpha_z}{\alpha_z +\tau_s}}  \xi^{-1/2} h_\mathrm{gas}
\end{equation}
\citep{YoudinLithwick2007}\footnote{The \citet{YoudinLithwick2007} study pertains to non-stratified disks. \Eq{hp-YL07} was suggested to account for stratification effects.} where $h_\mathrm{gas} =H_\mathrm{gas}/r$ and
\begin{equation}
    \xi = 1 +\frac{\tau_s(\Omega t_\mathrm{corr})^2}{\tau_s +\Omega t_\mathrm{corr}}.
    \label{eq:xi}
\end{equation}
In the limit of $\Omega t_\mathrm{corr}\ll 1$ or $\tau_s\ll (\Omega t_\mathrm{corr})^{-1}$, $\xi\approx1$ and the pebble aspect ratio reduces to $h_P =\sqrt{\alpha_z/(\alpha_z +\tau_s)} h_\mathrm{gas}$
\citep{DubrulleEtal1995}.

For small stopping times ($\tau_s=10^{-2}$; left panels) the SCA and the SEOM overlap. From a numerical perspective the SCA is preferable as it is computationally much less intensive than the SEOM-method. For $\tau_s=10^{-1}$ the distributions slightly differ, as can best be seen from the lower panels.  For $\tau_s=1$ particles the differences between the two methods amount to several tens of percents at $z=0$, while towards the tails of the distribution the relative difference is larger even. The SEOM, however, is in perfect agreement with the \citet{YoudinLithwick2007} theory on diffusive transport. The reason is that, like \citet{YoudinLithwick2007}, the SEOM accounts for the vertical oscillation (epicyclic motion) of particles, whereas the SCA does not. The SCA does not account for the pebble's inertia and also does not involve a correlation time. From \eq{hp-YL07} we deduce that the SCA becomes invalid for $\tau_s>(t_\mathrm{corr}\Omega)^{-1}$.

In the above, we assumed that $t_\mathrm{corr}\approx\Omega^{-1}$, which is applicable in the ideal limit of MRI-turbulence \citep{SanoEtal2004,JohansenEtal2006i,CarballidoEtal2011}. However, for non-ideal effects as ambipolar diffusion, the correlation time is expected to be longer \citep{BaiStone2011,ZhuEtal2015}. From \eq{hp-YL07} it is clear that pebbles will be more strongly stratified for large values of the turbulent correlation time $t_\mathrm{corr}$. In \fg{particle-distr}b a case with a 10 times longer correlation times (but still the same $\alpha_z=10^{-2}$ vertical diffusivity, implying larger, longer-lived but less vigorous eddies) is presented. Its vertical distribution is much narrower than the canonical $t_\mathrm{corr}=1\,\Omega^{-1}$ turbulence. A case with $t_\mathrm{corr}=10^{2}\,\Omega^{-1}$ is also shown in \fg{particle-distr}c. Clearly, a degeneracy between turbulent correlation time $t_\mathrm{corr}$, diffusivity ($\alpha_z$), and stopping time ($t_\mathrm{stop}$) is present.

\subsection{Vertical gradient in the diffusivity}
\label{sec:Dz}
As an application of a more convoluted model, we consider a vertically dependent diffusion. In terms of $\alpha_z$ we adopt
\begin{equation}
    \alpha_z(z)
    = \frac{\alpha_\mathrm{mid} +\alpha_\mathrm{surface}(z/2H_\mathrm{gas})^4}{1+(z/2H_\mathrm{gas})^4}
    \label{eq:alpha-z}
\end{equation}
where $\alpha_\mathrm{mid}$ is the diffusivity in the midplane and $\alpha_\mathrm{surface}$ is the diffusivity in the upper regions. Such layered accretion \citep{Gammie1996} when the turbulence is confined to the upper regions, although our parameterization in \eq{alpha-z} is completely arbitrary.  We choose $\alpha_\mathrm{surface}=0.1$ and $\alpha_\mathrm{mid}=10^{-3}$. This profile is plotted in \fg{diffz} by the red curve. We further choose $\tau_s=10^{-2}$, such that $\tau_s/\alpha_z>1$ in the midplane (indicating settling) and $\tau_s/\alpha_z<1$ in the upper regions (indicating coupling to the gas).

In \fg{diffz} the gray and black histograms show the normalized distribution of $z$ obtained with the SCA and SEOM methods, respectively. The methods give consistent result. Clearly, the high $|z|$ regions are sparsely sampled as the probability to find a particle at these heights is low. However, the fact that pebbles can be stirred to these heights at all may be surprising given the low $\alpha_\mathrm{mid}$. This is illustrated with the dark blue curve in \fg{diffz}, which plots $P(z,h_\mathrm{mid})$ where $h_\mathrm{mid}\approx\sqrt{\alpha_\mathrm{mid}/\tau_s}h_\mathrm{gas}\approx 0.3h_\mathrm{gas}$. In fact, pebbles at $|z|\gtrsim2 H_\mathrm{gas}$ follow the gas distribution (light blue curve). Altogether $P(z)$ can be approximated as the sum of two distributions.  The majority of the pebbles ($\approx$90\%) follow the midplane distributions (given by $h_\mathrm{mid}$), but about 10\% of the pebbles follow the distribution given by the gas scaleheight.

\subsection{Local replenishment of small grains?}
The ability of turbulence to stir $\sim$mm-sized pebbles from the midplane to many gas scaleheights may offer an explanation for the persistent presence of small, (sub)-micron size particles in the disk surface, as deduced from near-IR observations \citep[\eg][]{JuhaszEtal2010}. From a theoretical perspective, the presence of small particles is problematic as they should quickly coagulate among themselves and then settle to the disk midplane \citep{NakagawaEtal1986,TanakaEtal2005,DullemondDominik2005}. This implies that the grains are replenished. The most common theory postulates that the replenishment occurs in the disk midplane regions. Here grains are produced by high-velocity collisions among pebble-sized particles, which are subsequently transported (by diffusion) to the disk surface \citep{BirnstielEtal2010i}.  
However, this is a rather indirect route to replenish small grains.  First, it is doubtful if pebbles in the midplane will fragment; the gas may not be sufficiently turbulent.  Second, it takes grains a time $\sim$$1/\alpha_z \Omega$ to diffuse, which is rather long, again when $\alpha_z$ is small; these small grains may simply collide before reaching the surface \citep{KrijtCiesla2016}.

Alternatively, in a disk with a suitable diffusivity profile ($\alpha_z(z)$), it is possible to diffuse a small number of pebbles to the disk surface. There, due to the much stronger turbulent velocity field as compared to the midplane, collisions will undoubtedly be catastrophic. To cement these ideas, a coupled transport-collision/fragmentation model need to be considered.

\section{3D pebble accretion}
\label{sec:results}
Following Paper I we calculate the accretion efficiency ($\epsilon$) by conducting a series of N-body integrations to follow the trajectory of pebbles as they drift from orbits exterior to the planet to orbits interior to it. The pebble accretion efficiency $\epsilon$ is then found simply by counting the fraction of particles that settle to the planet. While Paper I investigated the role of the planet's eccentricity, we fix $e_p=0$ here and instead investigate the role of turbulence and the planet inclination (\se{inclination}). To this effect we let the pebble experience a stochastic motion in the vertical direction, as outlined in \eq{eom-all}. The initial vertical position $z_\mathrm{ini}$ is given by the steady-state distribution characterized by the pebble scaleheight $h_P$. The initial radial position is set by
\begin{equation}
    r_\mathrm{ini} = a_p +3b_\mathrm{shear} +\Delta r_\mathrm{syn}(\eta,\tau_s,N) +0.1r_\mathrm{Hill}
    \label{eq:rini}
\end{equation}
where $b_\mathrm{shear}=\tau_s^{1/3} r_\mathrm{Hill}$ is the impact parameter for pebble accretion in the Hill regime and $\Delta r_\mathrm{syn}(\tau_s,\eta,N)$ the distance a pebble drifts after $N$ synodical orbits.\footnote{We calculate $\Delta r_\mathrm{syn}$ from the equation
\begin{equation}
    \int_{r_p}^{r_p+\Delta r_\mathrm{syn}} \frac{\Omega_p-v_\phi(r')/r'}{v_r} \dd r'
    = 2\pi N
\end{equation}
where $v_r$ and $v_\phi$ are the radial and azimuthal drift velocities \citep[\eg][]{Weidenschilling1977}. Approximating the integral to second order in $\Delta r_\mathrm{syn}$, we obtain
\begin{equation}
    \Delta r_\mathrm{syn} = \frac{-\eta +\sqrt{\eta^2 +(2\pi N)4\eta\tau A_{\tau,\eta}}}{A_{\tau,\eta}} r_p
    \label{eq:delrsyn}
\end{equation}
where $A_{\tau\eta} = \frac{3}{2}(1 +\tau_s^2) -\eta$.
}
We choose $N=3$. When the pebble radius has drifted to a distance $r_p -r_\mathrm{Hill}$ we stop the calculation. It is then counted as a miss.

The fact that particles are only kicked in the $z$-direction, allows us to restrict the computations to a narrow ring. In contrast, when we would have considered the general case (turbulence operating in all dimensions), a much larger computational domain would be required because of the possibility of multiple encounters (similar to the eccentric case in Paper I). This complication is the key reason why we consider only turbulence in the vertical dimension.

\begin{figure*}[t]
    \sidecaption
    \includegraphics[width=0.7\textwidth]{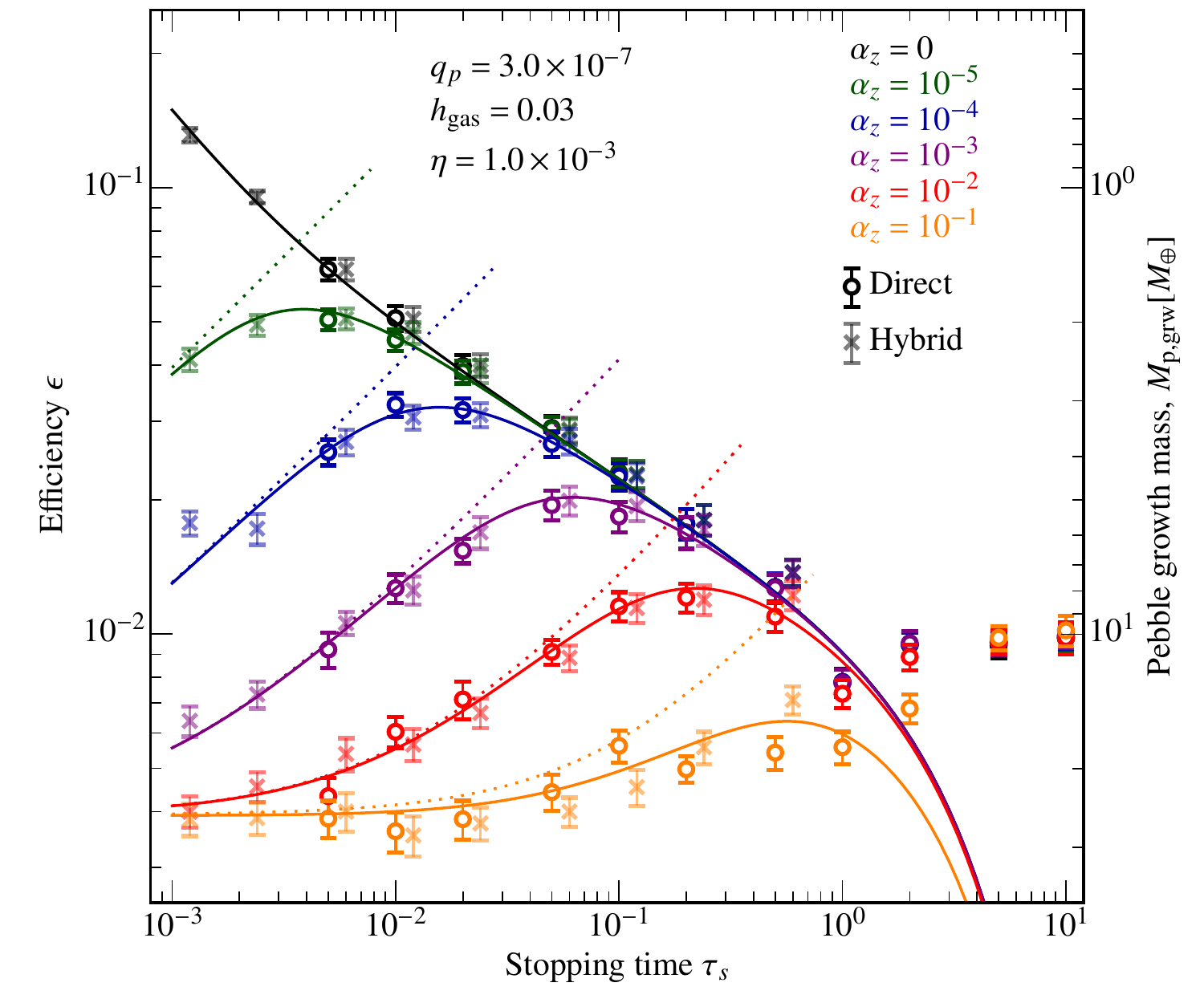}
    \caption{Pebble accretion efficiency vs dimensionless stopping time for several values of the vertical turbulence strength, parameterized by $\alpha_z$ (colors). Open symbols give $\epsilon$ obtained from directly integrating the stochastic equation of motion, while crosses give the results from the hybrid algorithm. Crosses are offset by 20\% to the right for clarity. Error bars correspond to the Poisson counting error on the number of hits. For $\alpha_z=0$ the triangles give $\varepsilon$ resulting from the local calculations and the black line gives our fit to the 2D-limit ($\epsilon_\mathrm{2d}$), which we obtained in Paper I. The dotted lines gives $\epsilon_\mathrm{3D}$ accounting just for the density correction \eq{eps-3D}. The colored solid lines give $\varepsilon$ accounting for all 3D/2D-effects (\eq{eps-combine}) as explained in the main text. The right $y$-axis converts $\varepsilon$ into the pebble growth mass -- the total amount of pebbles needed to $e$-fold the mass of the planet -- assuming a solar-mass star.
}
    \label{fig:standard}
\end{figure*}
Similar to Paper I, we express lengths in terms of the disk radius $r_p$ and times in units of $\Omega^{-1}$. The key parameters are:
\begin{itemize}
    \item $q_p = M_p/M_\star$ the planet-to-stellar mass ratio;
    \item $h_\mathrm{gas} = H_\mathrm{gas}/r$ the disk aspect ratio at the location of the planet;
    \item $\eta$, a measure of the radial drift velocity of the pebbles (\eq{eta});
    \item $\tau_s=t_\mathrm{stop}\Omega$, the dimensionless stopping time;
    \item $\alpha_z$, a proxy for the gas vertical diffusivity $D_{zz}$;
    \item $t_\mathrm{corr}$, the turbulent correlation time, which together with $\alpha_z$ determines the magnitude of the turbulent rms velocities, \eq{sigma-z}.
\end{itemize}
Analogous to Paper I, we consider two integration methods:
\begin{itemize}
    \item The SEOM, which integrates the particle velocity (\eq{eom-all});
    \item The hybrid method, which uses the SCA, but switches to the SEOM when the particle is in the vicinity of the planet. Here we take the criterion to switch to the SEOM to be $\sqrt{(\Delta x)^2 +(\Delta y)^2} < 2r_\mathrm{Hill}$, where $\Delta x$ and $\Delta y$ are the distances between the planet and pebble in the $x$ and $y$ Cartesian coordinates. Note the absence of the vertical position in this criterion. The reason is that for some parameter combinations (small planet mass and high $\alpha_z$) the vertical step size can easily become larger than the Hill radius in the SCA method.
\end{itemize}
As explained in Paper I, the SCA assumes that gas and particles are well coupled. It does not capture effects that take place on timescales $\Delta t$ less than $t_\mathrm{stop}+t_\mathrm{corr}$. On small $\Delta t$ the particle moves ballistically, while the SCA model keep exhibiting random walk behavior at all (time)scales. Within the Hill sphere, where the numerical timestep will become small, the particle trajectories are therefore incorrect. The strong fluctuations in velocity space also complicates the numerical integration.

Different from Paper I, we do not account for the ballistic regime. Ballistic encounters are encounters in which the pebble is not captured by the settling mechanism, but where accretion occurs by virtue of the pebble hitting the surface of the target. Computationally, we can easily distinguish between ballistic and settling encounters by assigning an arbitrary small physical size to the planet (while keeping its mass). All accretion then occurs through the settling mechanism.

\subsection{Standard model and analytical fits}
For the standard model we take $q_p=3\times10^{-7}$ (a $0.1\,M_\oplus$ mass planet for a solar-mass star), $\eta=10^{-3}$, $h_\mathrm{gas}=0.03$, $t_\mathrm{corr}=\Omega^{-1}$ and vary $\alpha_z$ and $\tau_s$. The choices for $\eta$ and $h_\mathrm{gas}$ approximately correspond to a disk location of 1\,au; both values will generally be higher in the outer disk. In \fg{standard} symbols give the mean value of $\varepsilon$ obtained from our numerical integrations, $\varepsilon=N_\mathrm{set}/N_\mathrm{tot}$, where out of $N_\mathrm{tot}$ integrations $N_\mathrm{set}$ pebbles settled to the planet.
Error bars correspond to the Poisson error, $\sqrt{N_\mathrm{set}}/N_\mathrm{tot}$. The pebble accretion efficiency can be converted into the pebble growth mass, $M_\mathrm{P,grw}$, defined as
\begin{equation}
    M_\mathrm{P,grw} = \frac{q_p M_\star}{\varepsilon}
    \label{eq:mP-grw}
\end{equation}
(values labeled on the right $y$-axis). This is the amount of pebbles needed to $e$-fold the planet's mass. Finally, solid curves gives our fit to the data, which will be discussed in the subsequent sections and summarized in \se{fit}.  The fitting expression is appropriate only for $\tau_s\lesssim1$.

Clearly, where $\varepsilon$ is higher, fewer integrations are needed to obtain a minimum signal-to-noise. In our integrations $N_\mathrm{tot}$ is not fixed, but different for each run in order to obtain a certain signal-to-noise ratio, which is determined by the number of settling encounters. Hence, most of the computational effort is spent in the runs where $\varepsilon$ is small. Results of the SEOM method are shown by the open circles, while the crosses (slightly offset) denote the results from the hybrid method. The two methods give consistent results (see also Paper I). For small $\tau_s$ the SEOM method becomes computationally inefficient as it takes the pebble a long time to drift to the interior disk and the integration becomes very stiff. The hybrid method removes this latter problem and is the method of choice for small $\tau_s$. Remarkably, the hybrid method gives acceptable results up to $\tau_s=0.5$.

Results for the non-turbulent (2D) limit ($\alpha_z=0$; black and gray symbols) were already discussed in Paper I. The efficiency decreases towards increasing particle stopping times -- $\tau_s=1$ particles are accreted at the lowest efficiency -- because the faster drift by the higher $\tau_s$ particles outweighs the larger linear cross section. Consequently, the pebble growth mass $M_\mathrm{P,grw}$ is large;  many pebbles are needed in order to grow the planet. The steepening of the slope that can be noticed at small $\tau_s$ is the result of a transition from the shear regime (velocities dominated by the Keplerian shear) at high $\tau_s$ to the headwind regime (velocities dominated by the gas sub-Keplerian motion, $\eta v_K$) at small $\tau_s$.

When $\alpha_z>0$ (colored symbols) pebbles are stirred to higher regions, reducing the local density of pebbles in the midplane. This reduces $\varepsilon$ with respect to the 2D case. As can be seen in \fg{standard} pebbles are most affected when they are small ($\tau_s\ll1$) and when the turbulence is strong (high $\alpha_z$), which is of course natural. Hence, in the 3D case there is a preferred pebble aerodynamic size where $\varepsilon(\tau_s)$ peaks, which occurs approximately at the point when the pebble accretion impact parameter equals the pebble scaleheight, \ie\ at the transition of the 2D and 3D regimes. For heavier particles $\varepsilon$ decreases because of more rapid radial drift, whereas for small particles $\varepsilon$ decreases because of a reduced local density. However, when $\alpha_z \gtrsim \tau_s$ the pebble scaleheight has reached that of the gas, $h_P\approx h_\mathrm{gas}$, and no further reduction is possible. As a result, the curves eventually converge when $\tau_s$ becomes very small, as is seen in \fg{standard} in the bottom-left corner.

In \citet{Ormel2017}, as well as Paper I, we derived that the pebble accretion efficiency in the 3D limit reads
\begin{equation}
    \varepsilon_\mathrm{3D}
    = A_3 \frac{q_p}{\eta h_P} f_\mathrm{set}^2
    \label{eq:eps-3D}
\end{equation}
where $A_3$ is a numerical constant, $h_P$ the pebble aspect ratio, and $f_\mathrm{set}$ -- the settling fraction -- a modulation factor which becomes less than unity when the settling criteria (slow encounters) is no longer fulfilled. The characteristic velocity beyond which settling encounters disappear is
\begin{equation}
    \label{eq:vast}
    v_\ast = \left(\frac{q_p}{\tau_s}\right)^{1/3} v_K
\end{equation}
(\citealt{OrmelKlahr2010}; Paper I). Qualitatively, when the approach velocity\footnote{the magnitude of the unperturbed relative velocity between planet and pebble.} $\Delta v \ll v_\ast$ settling is fully operational ($f_\mathrm{set}=1$), while for $\Delta v \gg v_\ast$ settling (and therefore pebble accretion) are no longer viable ($f_\mathrm{set}=0$). Quantitatively, the settling modulation function is fitted empirically by an exponential function \citep{OrmelKlahr2010,VisserOrmel2016}. In Paper I we adopted
\begin{equation}
    f_\mathrm{set,I} = \exp \left[ -a_\mathrm{set} \left( \frac{\Delta v}{v_\ast} \right)^2 \right].
    \label{eq:fsetI}
\end{equation}
with $a_\mathrm{set}=0.5$. The subscript ``I'' indicates that this expression is valid for a laminar disk (Paper I). We will refine it below (\se{turbulence}) accounting for a turbulence velocity field.

\begin{figure*}[t]
    \centering
    \includegraphics[width=0.85\textwidth]{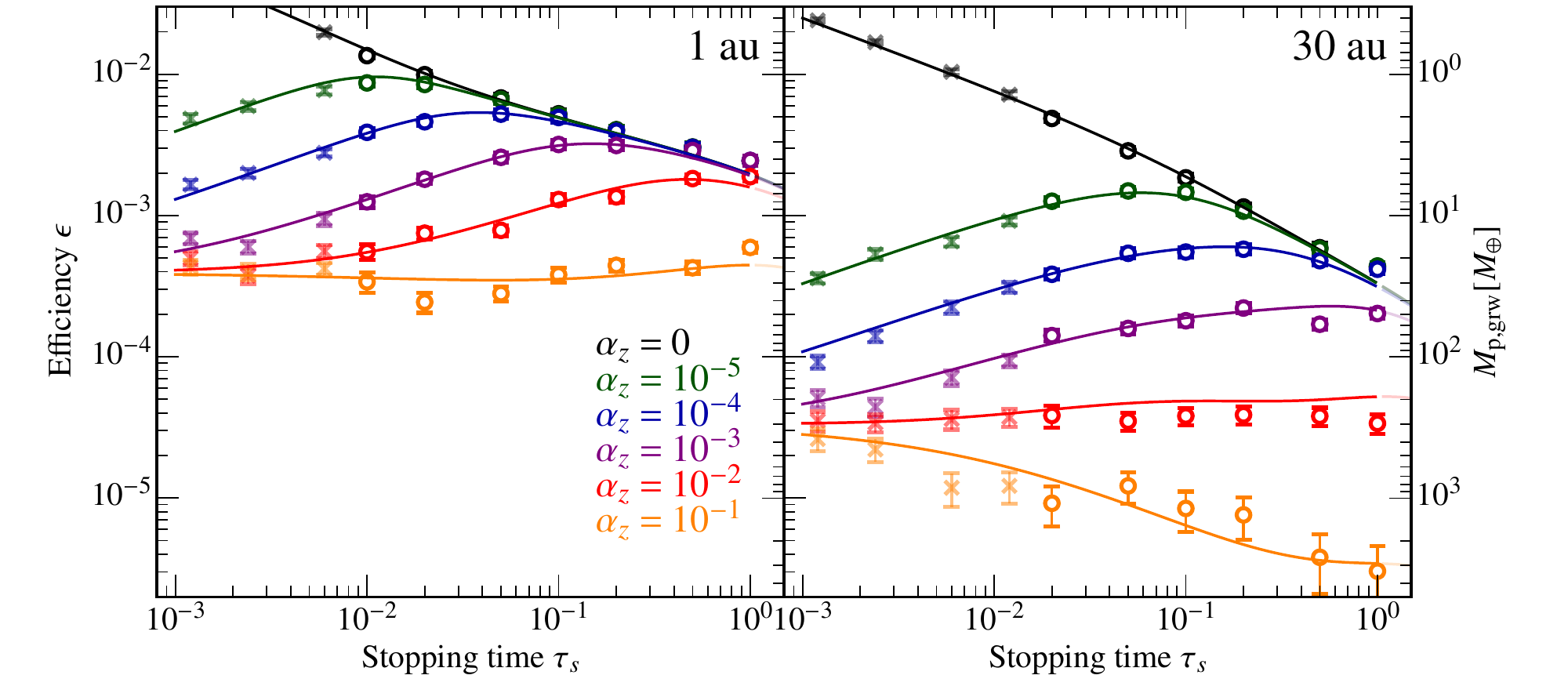}
    \caption{Efficiency of pebble accretion of a $0.01\,M_\oplus$ planet ($q_p=3\times10^{-8}$) at 1 au (left) and at 30 au (right). At 1 au $h_\mathrm{gas}=0.03$ and $\eta=10^{-3}$, while at 30 au we take $h_\mathrm{gas}=0.07$ and $\eta=5\times10^{-3}$. Right axis gives the planet growth mass. At 30 au turbulence more significantly affects the pebble accretion efficiency.}
\label{fig:outer}
\end{figure*}
The reduction of $\varepsilon$ by $f_\mathrm{set}$ enters quadratically in 3D because the cross section is two-dimensional. By virtue of the rather high planet mass, $f_\mathrm{set}$ nevertheless evaluates to unity for most runs in \fg{standard}. In \fg{standard} the dotted lines give \eq{eps-3D}, where we took $f_\mathrm{set}=1$ and evaluated $h_\mathrm{peb}$ according to \eq{hp-YL07}. With $A_3=0.39$ this matches the numerical results well for small $\tau_s$. At high $\tau_s$, \eq{eps-3D} clearly overestimates the pebble accretion efficiency; the settling efficiency is then given by its planar limit.

Only for $\alpha_z=0.1$ tend the efficiencies to lie below the expression given by \eq{eps-3D}. The reason is that now the pebble velocity becomes dominated by turbulent motions since $\alpha_z^{1/2}h_\mathrm{gas} \gtrsim \eta$, which suppresses accretion through settling as encounters become too fast for settling, $f_\mathrm{set}<1$, because of a high turbulent velocity. We present a model to include for turbulence effect in $f_\mathrm{set}$ in \se{turbulence}.

\subsection{Pebble accretion in the outer disk}
For planets on circular orbits, the pebble accretion efficiency is fully determined by the five dimensionless parameters $q_p$, $h_\mathrm{gas}$, $\eta$, $\tau_s$, and $\alpha_z$. In the outer disk, the aspect ratio $h_\mathrm{gas}$ and (as a consequence) $\eta$ are usually higher, resulting in much lower efficiencies. This is illustrated in \fg{outer}, which shows $\varepsilon(\tau_s,\alpha_z)$ for the inner disk (left) and outer disk (right) for a $q_p=3\times10^{-8}$ planet ($0.01\,M_\oplus$ for a solar-type star).  For the outer disk run we have increased $h_\mathrm{gas}$ and $\eta$ by factors of $\approx$2 and $\approx$5, respectively. For a standard passively irradiated disk model this would correspond to an increase by a factor 30 in $r_p$; \eg we contrast the situation at 1 au (left) with 30 au (right).

Efficiencies in the outer disk are always lower -- in particular, the 3D efficiencies. In addition, the transition to the 2D limit occurs at a longer stopping time. Both effects are caused by the larger gas scaleheight. A further consequence of a larger $h_\mathrm{gas}$ is that turbulence velocities become higher compared to the critical threshold $v_\ast$ (both velocities are lower in the outer disk, but whereas $\sigma_z\propto c_s \propto h_\mathrm{gas}v_K$,  $v_\ast \propto v_K$). Consequently, for the 30 au run, settling already fails for $\alpha=10^{-2}$, which manifests itself by the flattening and decrease of the curves.

It has been suggested that pebble accretion is an effective mechanism to grow planets in the outer disk \citep{OrmelKlahr2010,LambrechtsJohansen2012,BitschEtal2015,JohansenLambrechts2017}. Compared to planetesimal accretion the key advantage is that capture radii are large due to the large Hill radii, whereas planetesimals suffer from scattering -- a negative feedback to the growth of planets \citep{KobayashiEtal2010}. Nevertheless, as \fg{outer} illustrates, the efficiency of pebble accretion in the outer disk is lower than in the inner disk. For example, growing planets in strongly turbulent ($\alpha_z>10^{-2}$) disks may require hundreds, if not thousands, of Earth masses in pebbles -- numbers that seem rather large in the light of recent ALMA observations \citep{AnsdellEtal2017,MiotelloEtal2017}. Still, invoking (a combination of) low turbulence (as suggested by HL tau; \citealt{PinteEtal2016}), pressure bumps \citep{PinillaEtal2012}, or massive disks \citep{BitschEtal2018} there is enough leeway to grow planets through pebble accretion in the outer disks. But from an efficiency perspective, it is more conducive to grow them in the inner disk.

\begin{figure}[t]
    \includegraphics[width=0.5\textwidth]{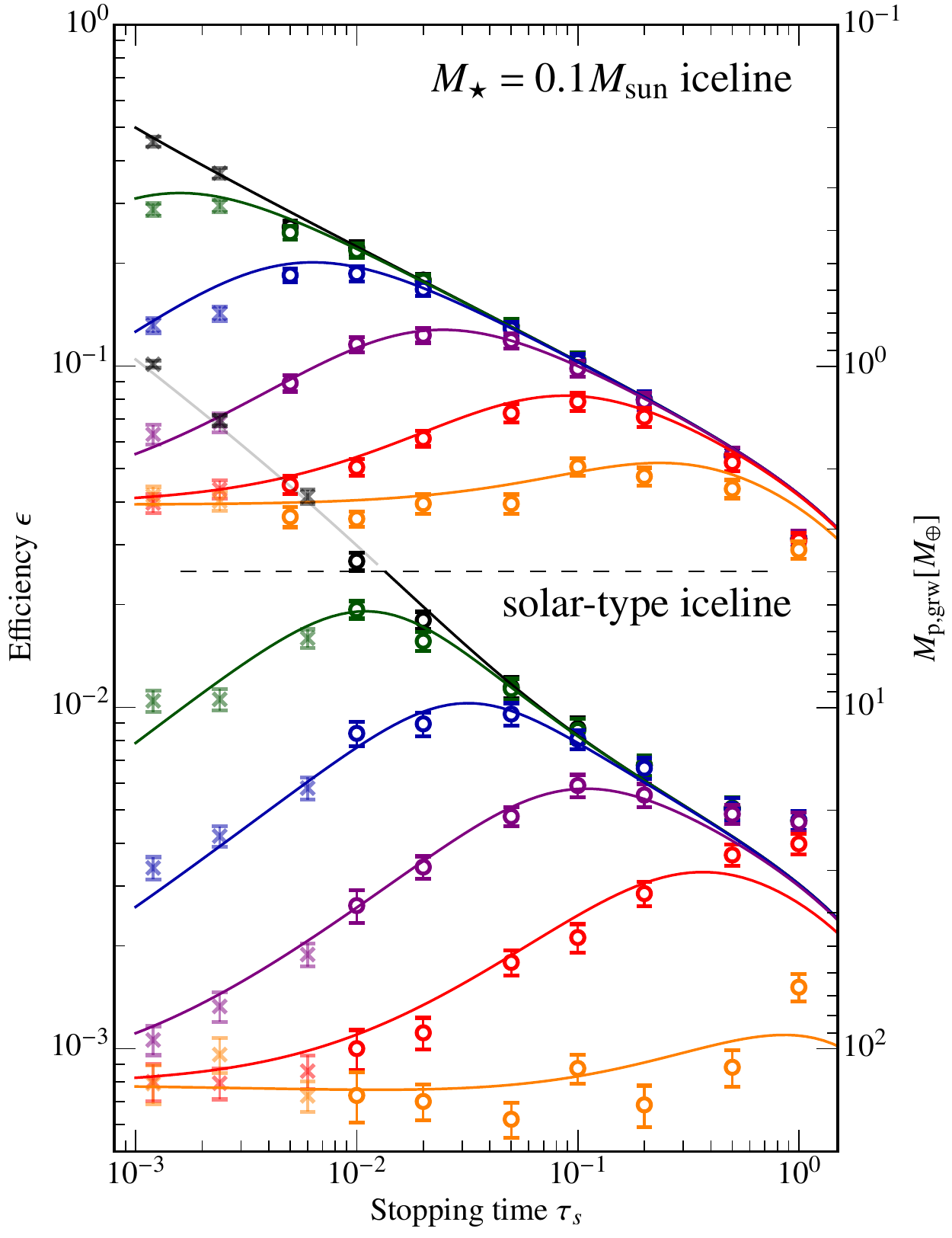}
    \caption{Pebble accretion efficiency for a 0.1 $M_\oplus$ planet at the location of the H$_2$O iceline. Top: Iceline around a 0.1 $M_\star$ mass star ($h_\mathrm{gas}=0.03$, $\eta=10^{-3}$; $q_p=3\times10^{-6}$). Bottom: iceline of a solar mass star ($h_\mathrm{gas}=0.05$, $\eta=3\times10^{-3}$, and $q_p=3\times10^{-7}$).}
    \label{fig:qpvar}
\end{figure}
\begin{figure*}[t]
    \centering
    \includegraphics[width=0.85\textwidth]{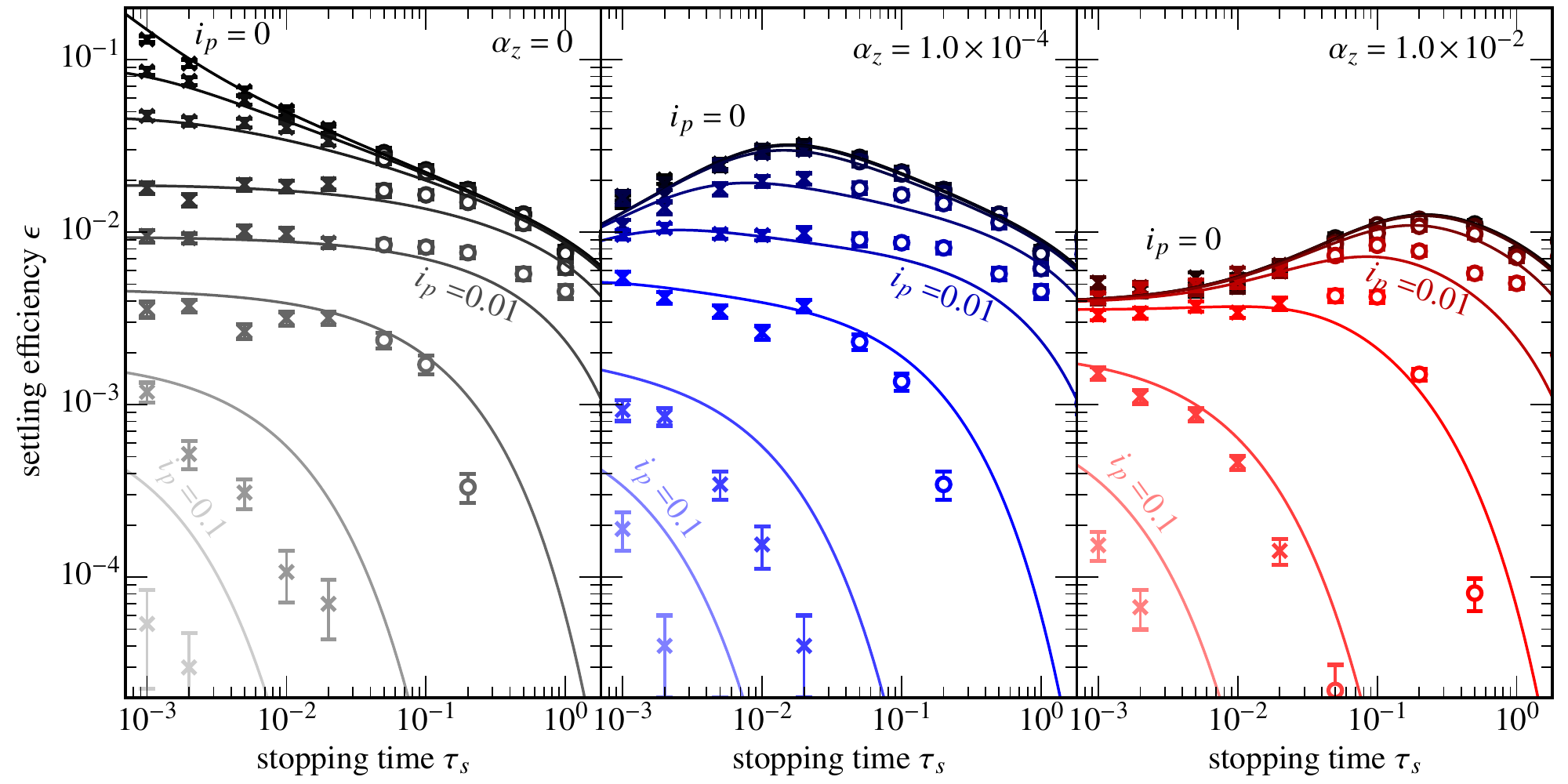}
    \caption{Planet inclination dependence on pebble accretion efficiency. We plot $\varepsilon$ for our standard parameters ($q_p=3\times10^{-7}$, $\eta=10^{-3}$ and $h_\textrm{gas}=0.03$) as function of $\tau_s$ (x-axis), $\alpha_z$ (panels) and planet inclination $i_p$ (colors). Curves give our fit (\se{fit}). The plotted planet inclination values are $i_p=0$ (black, top), $10^{-3}$, $2\times10^{-3}$, $5\times10^{-3}$, $10^{-2}$, $2\times10^{-2}$, $5\times10^{-2}$ and $0.1$ radians. Many of the low $i_p$ points and curves overlap. }
\label{fig:inclin}
\end{figure*}
\subsection{Efficiency at H$_2$O iceline for different stellar mass}
There have been a number of studies that argue for the H$_2$O iceline as a preferential site for the formation of the (first) generation of planetesimals and planetary embryos \citep{CuzziZahnle2004,RosJohansen2013,IdaGuillot2016,BanzattiEtal2015,SchoonenbergOrmel2017,DrazkowskaAlibert2017}. In many of these works, the abundance of ices increases just outside the iceline by condensation of H$_2$O vapor that diffused back over the iceline. In addition the surface density increases through a ``traffic jam'' effect when evaporating ice boulders liberate much smaller grains. For these reasons it is worthwhile to consider the efficiency of pebble accretion at the snowline. However, the snowline locations vary with stellar mass. In disks of lower mass stars the snowline will be much closer in, where the disk aspect ratio is likely to be smaller.

In \Fg{qpvar} the pebble accretion efficiency is plotted at the H$_2$O iceline, contrasting a solar type star (bottom) with a late M-star (top). The planet mass is $0.1\,M_\oplus$ in both cases. Two effects conspire to render efficiencies much higher for icelines around low-mass stars. First, the aspect ratio and $\eta$ are lower because the iceline lies further in. Second, a planet(esimal) of the same mass around a lower mass star will have a higher mass ratio ($q_p$). The pebble accretion cross section is then larger because of the reduced Keplerian shear and headwind velocities. \Fg{qpvar}b illustrates the effect for a $0.1\,M_\odot$ M-star for (the same) $0.1\,M_\oplus$ mass planet with $h_\mathrm{gas}=0.03$ and $\eta=10^{-3}$; i.e., $q_p$ is higher by a factor of 10, $h_\mathrm{gas}$ lower by a factor of 1.6, and $\eta$ lower by a factor 3 compared to the solar-type star. Clearly, M-star pebble accretion efficiencies are both higher and are less sensitive to variations in $\tau_s$ and $\alpha_z$. Therefore, (late type) M-stars can efficiently convert their pebble-sized building blocks into planetary systems. These finding confirm our earlier analytical estimates on a pebble formation origin of the TRAPPIST-1 system \citep{OrmelEtal2017}.

\subsection{Dichotomy of the solar system}
It was argued by \citet{MorbidelliEtal2015} that pebble accretion is more efficient beyond the H$_2$O iceline because icy pebbles, being less prone to collisional fragmentation, are larger (higher $\tau_s$). Specifically, \citet{MorbidelliEtal2015} adopted $\tau_s=10^{-1.5}$ for icy pebbles outside the iceline and $\tau_s=10^{-2.5}$ for silicate pebbles (just) interior to the iceline and adopted $\alpha_z=10^{-3}$.  The curve corresponding to these parameters is the purple curve in the bottom panel of \fg{qpvar}.  The positive slope indeed indicates that larger (icy) pebbles have a higher $\varepsilon$ than the smaller (silicate) pebbles, a situation that generally applies to the 3D limit (\eq{eps-3D} where the pebble aspect ratio $h_P$ decreases with higher $\tau_s$.
\citet{MorbidelliEtal2015} concluded that embryos just outside the H$_2$O iceline will outcompete inner embryos\footnote{Apart from a higher $\varepsilon$ another reason is that the icy pebble flux is larger than the silicate pebble flux, due to evaporation of H$_2$O.}, arguing that the great dichotomy of the solar system -- small Mars next to big Jupiter -- is well explained under a pebble accretion scenario.

There is a corollary to this hypothesis, not (explicitly) addressed by \citet{MorbidelliEtal2015}. As \fg{qpvar} shows the efficiencies of the $\tau_s=10^{-1.5}$ pebbles ($\varepsilon\approx0.4\%$) are very small: $>$99\% of the pebbles drift past the planet to enrich the inner solar system.  This means that the formation of Jupiter's core is associated with hundreds of Earth masses in pebbles drifting into the terrestrial planet region.  Evidently, in the scenario outlined by \citet{MorbidelliEtal2015} most pebbles could not have been accreted by the bodies in the inner solar system but must have ended up in the young Sun. This puts a constraint on the structure of the inner disk, \ie it had to be transparent to pebble drift. No dense planetesimals belts or long-lived pressure maxima (which could trigger formation of super-Earths; \citealt{ChatterjeeTan2014}) could have existed in the inner solar system during Jupiter's formation.

\subsection{Inclined planets}
\label{sec:inclination}
When planet(esimal)s move on orbits inclined with respect to the disk, the pebble accretion efficiency can be significantly reduced \citep{JohansenEtal2015,LevisonEtal2015}. Planets of high inclination $i_p$, such that $i_p>h_P$, only interact with pebbles over a fraction $\sim$$i_p/h_P$ of their orbits. Inclined planets, therefore, have a similar effect on the settling efficiencies as turbulence: the higher the inclination, the less it interacts with the pebbles. In addition, inclined planets encounter pebbles at an additional velocity ($\sim$$i_pv_K$), which suppresses accretion once it exceeds $v^\ast$.

These effects are illustrated in \fg{inclin}, where $\varepsilon$ is plotted as function of inclination $i_p$ (curves) for the laminar disk ($\alpha_z=0$; all pebbles in the midplane), $\alpha_z=10^{-4}$, and $\alpha_z=10^{-2}$. In general, the higher the inclination, the lower the accretion efficiencies. For $\alpha_z=0$ inclination effects already become visible for $i_p\sim10^{-3}$, whereas for $\alpha_z=10^{-2}$ the curves only diverge for $i_p>10^{-2}$ (inclinations are given in radians). Accretion of large $\tau_s$-particles in particular are suppressed because of the increase in the approach velocity and the decrease in $f_\mathrm{set}$. The planet moves too fast through the pebble plane.

Since the planet's inclination has a similar effect on $\varepsilon$ as turbulent stirring of pebbles -- both reduce the amount of interaction between the two-components -- an effective scaleheight may be defined as $h_\mathrm{eff} \simeq i_p +h_P$. A more precise estimate is obtained by averaging the pebble density over the phase of the planet's orbit
\begin{equation}
    \frac{1}{\sqrt{2\pi}h_\mathrm{eff}} = \int_0^{2\pi} \mathcal{N}(i_p\sin t, h_P) \frac{\dd t}{2\pi}
    \label{eq:rhoz}
\end{equation}
where $\mathcal{N}(z,h_P)$ is the normal distribution with standard deviation $h_P$ and $t$ the phase (mean anomaly). In the limit of $i_p\ll h_P$, $\mathcal{N}(0,h_P)=1/\sqrt{2\pi}h_P$ and $h_\mathrm{eff}$ is equal to the pebble scaleheight. More generally, the formal solution to \eq{rhoz} reads
\begin{equation}
    h_\mathrm{eff}
    = h_p \frac{\exp\left[ i_p^2/4h_P^2 \right]}{I_0(i_p^2/4h_P^2)}
    \label{eq:hs-Bessel}
\end{equation}
where $I_0(x)$ is the modified Bessel function of the first kind. For practical purposes \eq{hs-Bessel} may be approximated as
\begin{equation}
    \label{eq:hs-app}
    h_\mathrm{eff} \approx \sqrt{h_P^2 +\frac{\pi i_p^2}{2} \left( 1-\exp\left[ -i_p/2h_P \right] \right)}
\end{equation}
In the limit $i_p\gg h_P$, $h_\mathrm{eff}\simeq \sqrt{\pi/2} i_p \approx 1.25 i_p$, implying that our first guess ($h_\mathrm{eff}=i_p$) is off by 25\%. The reason is that the inclined planet spends most of its time near its end points, whereas it moves quickly through the midplane regions.

\subsection{Role of turbulence -- refinement of $f_\mathrm{set}$}
\label{sec:turbulence}
We refine the expression for $f_\mathrm{set}$ (\eq{fsetI}) accounting for a, possibly anisotropic, turbulent velocity field. Let $\Delta \bm{v}$ be the non-turbulent component of the approach velocity with its component, $\Delta v_i$, pointing in the $i^\mathrm{th}$ direction. Concerning the planar direction, we already obtained $\Delta v$ in Paper I, which we now relabel $\Delta v_y$:
\begin{equation}
    \label{eq:vy}
    \Delta v_y = \max \left( v_\mathrm{cir}, v_\mathrm{ecc} \right)
\end{equation}
where
\begin{equation}
    \label{eq:vcir}
    \frac{v_\mathrm{cir}}{v_K} =
    \frac{\eta}{ 1 +a_\mathrm{cir} q_p\tau_s/\eta^3} +a_\mathrm{sh} (q_p \tau_s)^{1/3}
\end{equation}
is the approach velocity in the circular limit. It combines the headwind (approach velocity dominated by $\eta v_K$) and the shear (approach velocity determined by the Keplerian shear velocity) regimes. In addition
\begin{equation}
    v_\mathrm{ecc} = a_e e_p v_K
    \label{eq:vecc}
\end{equation}
is the eccentric velocity. In the above formulae $a_\mathrm{cir}$, $a_e$ and $a_\mathrm{sh}$ are all fit constants (see \Tb{fitpars}).

Similarly, for the vertical approach velocity we have
\begin{equation}
    \Delta v_z = a_i i_p v_K
    \label{eq:vz}
\end{equation}
form the planet's inclination. The order of unity prefactor $a_i$ is again obtained numerically.

We assume a tri-axial Gaussian velocity distribution of width $\bm{\sigma}_P$, centered on $\Delta\bm{v}$.
Explicitly, for component $i$ the approach velocity $v_i$ is normally distributed
\begin{equation}
    P(v_i) = \frac{1}{\sigma_{\mathrm{P},i}\sqrt{2\pi}} \exp \left[ - \frac{(v_i -\Delta v_i)^2}{2\sigma_{\mathrm{P},i}^2} \right]
    \label{eq:Pvi}
\end{equation}
where $\sigma_{\mathrm{P},i}$ is the turbulent rms velocity in direction $i$. Following \citet{YoudinLithwick2007} we take
\begin{equation}
    \sigma_{\mathrm{P},i}
    = \sqrt{\frac{\alpha_i}{\Omega t_\mathrm{corr}+\tau_s}} \xi^{-1/2} hv_K
    \label{eq:sigmaP}
\end{equation}
for the pebble rms velocity where $\xi$ is defined in \eq{xi}.  From \eq{fsetI} we can write for the accretion probability
\begin{equation}
    \label{eq:facc}
    f_\mathrm{acc} = \exp\left[-a_\mathrm{set} \frac{v_x^2 +v_y^2 +v_z^2}{v_\ast^2} \right].
\end{equation}
The new, distribution-averaged $f_\mathrm{set}$ is then obtain by integration over the velocity distribution:
\begin{multline}
    \label{eq:fset}
    f_\mathrm{set}
    = \int f_\mathrm{acc} P(v_x) P(v_y) P(v_z) dv_x dv_y \dd v_z \\
    = \prod_i \exp \left[ -a_\mathrm{set} \frac{\Delta v_i^2}{v_\ast^2 +a_\mathrm{turb}\sigma_{\mathrm{P},i}^2} \right]  \frac{v_\ast}{\sqrt{(v_\ast^2 +a_\mathrm{turb}\sigma_{\mathrm{P},i}^2)}}.
\end{multline}
Formally, the integration gives $a_\mathrm{turb}=2a_\mathrm{set}$, but we relax this constant in order to obtain the best fit to the simulated data.

\Eq{fset} features the following limits:
\begin{enumerate}
    \item $\sigma_{\mathrm{P},x},\sigma_{\mathrm{P},y},\sigma_{\mathrm{P},z} \ll v_\ast$. Turbulence is unimportant and the laminar form of $f_\mathrm{set}$ is retrieved, \eq{fsetI}.
    \item Isotropic turbulence, $\sigma_{\mathrm{P},x}=\sigma_{\mathrm{P},y}=\sigma_{\mathrm{P},z} = \sigma_P$. This simplifies \eq{fset} to
    \begin{equation}
        \label{eq:fset-iso}
        f_\mathrm{set}
        = \exp \left[ -a_\mathrm{set} \frac{\Delta v^2}{v_\ast^2 +a_\mathrm{turb}\sigma_P^2} \right] \frac{v_\ast^3}{(v_\ast^2 +a_\mathrm{turb}\sigma_P^2)^{3/2}}
    \end{equation}
\item $\sigma_{\mathrm{P},x}=\sigma_{\mathrm{P},y}=0$ with turbulence only operating in the vertical direction, which is (by construction) the case considered in this paper and may be applicable to the vertical shear instability \citep{StollEtal2017}. Hence, we have used here
        \begin{equation}
            f_\mathrm{set} = \exp\left[ -a_\mathrm{set} \left( \frac{\Delta v_y^2}{v_\ast^2} +\frac{\Delta v_z^2}{v_\ast^2 +a_\mathrm{turb}\sigma_{\mathrm{P},z}^2} \right) \right] \frac{v_\ast}{\sqrt{v_\ast^2 +a_\mathrm{turb}\sigma_{\mathrm{P},z}^2}}.
        \end{equation}
    \item $\sigma_{\mathrm{P},x}\simeq\sigma_{\mathrm{P},y}\simeq\sigma_{\mathrm{P},z}\simeq\sigma_P \gg v_\ast+\Delta v$, the turbulence-dominant limit. In this case
        \begin{equation}
            f_\mathrm{set} = \left( \frac{v_\ast}{a_\mathrm{turb}^{1/2} \sigma_P} \right)^3.
        \end{equation}
        Turbulence reduces the accretion, but not exponentially. In a turbulence-dominated velocity field there is always a fraction of particles with velocities low enough to accrete by settling.
\end{enumerate}

\begin{table}
    \caption{Breakdown of the expressions and parameters involved in the pebble accretion efficiency.}
    \centering
    \small
    \begin{tabular}{llp{5cm}}
    \hline
    Expr.  & Dependence & Description and reference \\
                & or definition    & \\
    \hline
    $\varepsilon_\mathrm{set}$  & $\varepsilon_\mathrm{2D}$, $\varepsilon_\mathrm{3D}$  & settling efficiency,                        \eq{eps-combine} \\
    $\varepsilon_\mathrm{2D}$   & $A_2$, $\Delta v_y$, $f_\mathrm{set}$                 & settling efficiency in 2D limit,   \eq{eps-2D-2}\\
    $\varepsilon_\mathrm{3D}$   & $A_3$, $h_\mathrm{eff}$, $f_\mathrm{set}$             & settling efficiency in 3D limit, \eq{eps-3D-2}\\
    $f_\mathrm{set}$  & $\Delta \bm{v}$, $\bm{\sigma}_P$, $v_\ast$, $a_\mathrm{turb}$     & settling fraction,                          \eq{fset}         \\
    $\Delta v_y$                & $v_\mathrm{cir}$, $v_\mathrm{ecc}$                    & azimuthal approach velocity,                \eq{vy} \\
    $\Delta v_z$                & $a_i$                                                 & vertical approach velocity,                 \eq{vz} \\
    $v_\ast$                    & $(q_p/\tau_s)^{1/3} v_K$                              & critical settling velocity                          \\
    $v_\mathrm{cir}$            & $a_\mathrm{cir}$, $a_\mathrm{sh}$                     & circular velocity,                          \eq{vcir} \\
    $v_\mathrm{ecc}$            & $a_e$                                                 & eccentric velocity,                         \eq{vecc} \\
    $h_\mathrm{eff}$            & $i_p$, $h_P$                                          & effective aspect ratio,                     \eq{hs-app} \\
    $h_P$                       & $\alpha_z$, $t_\mathrm{corr}$                         & pebble aspect ratio,                        \eq{hp-YL07} \\
    $A_2$                       & 0.32                                                  & fit constant \\
    $A_3$                       & 0.39                                                  & fit constant \\
    $a_\mathrm{cir}$            & 5.7                                                   & fit constant \\
    $a_\mathrm{e}$              & 0.76                                                  & fit constant \\
    $a_\mathrm{i}$              & 0.68                                                  & fit constant \\
    $a_\mathrm{set}$            & 0.5                                                   & fit constant \\
    $a_\mathrm{sh}$             & 0.52                                                  & fit constant \\
    $a_\mathrm{turb}$           & 0.33                                                  & fit constant \\
    $\alpha_z$                  &                                                       & turbulent diffusivity \\
    $\eta$                      &                                                       & disk radial pressure gradient parameter, \eq{eta} \\
    $\tau_s$                    & $t_\mathrm{stop}\Omega$                               & dimensionless stopping time \\
    $\bm{\sigma}_P$             & $(\sigma_x,\sigma_y,\sigma_z)$                        & pebble rms velocity (\eq{sigmaP}) \\
    $e_p$                       &                                                       & planet eccentricity \\
    $h_\mathrm{gas}$            & $H_\mathrm{gas}/r$                                   & disk aspect ratio \\
    $i_p$                       &                                                       & planet inclination \\
    $q_p$                       & $M_\mathrm{pl}/M_\star$                              & planet-to-star mass ratio \\
    \hline
    \end{tabular}
    \label{tab:fitpars}
\end{table}

\subsection{Summary of the pebble accretion efficiency fit}
\label{sec:fit}
We provide an executive summary on how to generally obtain $\varepsilon$. Quantities involving the recipe and corresponding numerical constants are given in \Tb{fitpars}.

The settling efficiencies in the 2D and 3D limits read
\begin{subequations}
\label{eq:eps-final}
\begin{equation}
    \varepsilon_\mathrm{2D,set} = \frac{A_2}{\eta} \sqrt{\frac{q_p}{\tau_s}\frac{\Delta v}{v_K}} f_\mathrm{set}
    \label{eq:eps-2D-2}
\end{equation}
\begin{equation}
    \varepsilon_\mathrm{3D,set} = A_3 \frac{q_p}{\eta h_\mathrm{eff}}
    f_\mathrm{set}^2
    \label{eq:eps-3D-2}
\end{equation}
\end{subequations}
Apart from the $f_\mathrm{set}$ term, the 3D expression is independent of the pebble-particle relative velocity; a characteristic feature of pebble accretion in the 3D limit (\citealt{Ormel2017}; Paper I). The effective scaleheight, $h_\mathrm{eff}$, \eq{hs-app}, accounts for the reduced interaction between pebble and planetesimal by either turbulent stirring of pebbles or planetesimal inclination.
The relative motion or pebble approach velocity (barring turbulence) $\Delta v$ is given in \eq{vy} for the planar motion and \eq{vz} for the vertical motion due to planet inclination.

The other velocity-dependent term is the settling fraction $f_\mathrm{set}$, which becomes important when either laminar or turbulent velocities exceed the critical settling velocity $v_\ast = (q_p/\tau_s)^{1/3} v_K$. In the laminar case ($\sigma=0$) $f_\mathrm{set}$ is given by \eq{fsetI}, while in the turbulent case ($\sigma\neq0$) it is given by \eq{fset}. When $v_\ast$ significantly exceeds $\sigma+\Delta v$ the settling fraction evaluates to unity.

Finally, we find that the 3D (this Paper) and 2D (Paper I) can be combined as
\begin{equation}
    \varepsilon_\mathrm{set}
    = \left( \varepsilon_\mathrm{2d,set}^{-2} +\varepsilon_\mathrm{3D,set}^{-2} \right)^{-1/2}.
    \label{eq:eps-combine}
\end{equation}
which ensures a smooth transition. All curves shown in \fgss{standard}{xu} follow this recipe.

When $f_\mathrm{set}\ll1$ (\eg $\Delta v \gg v_\ast$ or $\sigma \gg v_\ast$) the ballistic regime (see Paper I) takes over. In Paper I we found that the combined efficiency may be given
\begin{equation}
    \varepsilon = f_\mathrm{set}\varepsilon_\mathrm{set} +(1-f_\mathrm{set})\varepsilon_\mathrm{bal}
    \label{eq:eps-tot}
\end{equation}
However, in the 3D limit, growth through ballistic interactions (gravitational focusing or geometric sweepup) is quite slow as pebble accretion cross section are generally much larger than their ballistic counterparts.

\subsection{Neglected effects}
We list several neglected effects, which may change $\varepsilon$:
\begin{enumerate}
    \item Aerodynamical deflection. For planetesimals, we have not accounted for the change in $\bm{v}_\mathrm{gas}$ in the vicinity of the planetesimals and correspondingly ignored any aerodynamic deflection \citep{VisserOrmel2016}. This reduction, however, only becomes important when particles are very small, $t_\mathrm{stop}<R/v_\mathrm{hw}$, and only when the encounters operate in the ballistic regime. Under these conditions, turbulence may in fact overcome the aerodynamic barrier \citep{HomannEtal2016}.
    \item Pre-planetary atmospheres. We have not accounted for the effects of the early primordial atmosphere that forms around massive bodies in gaseous disks. As a rule of thumb, this affects the density and flow pattern out to a Bondi radius, $R_b = GM_p/c_s^2$. Planets as massive as the thermal mass ($h^3M_\star$) will have such extensive envelopes that our constant-density assumption will break down. A complex flow structure may prevent small particles from accreting to the planet \citep{Ormel2013}, perhaps after their evaporation \citep{BrouwersEtal2018,Chambers2017}.
    \item Pressure maxima and resonances. Such massive planets also affect the pressure profile of the disk, resulting in a pressure maximum at a distance $\sim$$H_\mathrm{gas}$ from the planet. For this pebble isolation mass, accretion will terminate \citep{LambrechtsEtal2014,BitschEtal2018i}. Similarly, $\tau_s>1$ pebbles can be stopped at resonant location \citep{WeidenschillingDavis1985,PicognaEtal2018}.
    \item Gas radial flow. In viscous disks, gas flows radial at a velocity $\sim$$-\nu/r$ and adds to the drift velocity of pebbles. This effect hence starts to dominate radial drift motions for $\alpha_\nu > \tau_s$. It is straightforward to adjust expressions for $\varepsilon$ (see, \eg \citealt{IdaEtal2016}). Similarly, the drift velocity depends on vertical position, $v_r(z)$, because $\tau_s$ increases towards the disk surface. We can correct $\varepsilon$ accordingly, \ie by using a vertically averaged $v_r$ \citep{TakeuchiLin2002,KanagawaEtal2017}.
    \item By design, the simulations of this work only considered turbulence in the vertical direction -- a simplification that allowed us to carry out a numerical parameter study in a controlled way.  By modeling vertical turbulence we fully account for the diffusion effect on $\varepsilon$ -- \ie the reduction of the midplane pebble density. This will be unaffected by addition of planar turbulence.  However, the turbulent velocity effect -- \ie the reduction of $\varepsilon$ because of too fast encounters -- is determined by all velocity components. In particular, situation where $\sigma_x,\sigma_y \gg \sigma_z$ are not covered by our integration. Nevertheless, in \se{turbulence} we formulated a general recipe for $f_\mathrm{set}$ in a anisotropic velocity field.
\end{enumerate}

\begin{table}
    \caption{Collected efficiency expressions used in recent studies.}
    \centering
    \small
\begin{tabular}{llll}
    \hline
    limit                       & 2D/shear                      & 3D                & $i=e/2\gg h_P$\\
    Leading term $\varepsilon$  & $q_p^{2/3}/\eta\tau_s^{1/3}$  & $q_p/\eta h_P$    & $q_p/\eta i_P$        \\
    \hline \\[-2ex]
    This work/Paper I                               & \textbf{0.23} & \textbf{0.39} & \textbf{0.31}\\
    \citet{Chambers2014}\tablefootmark{a}           & 0.31      & 0.5       & 0.5 \\
    \citet{MorbidelliEtal2015}\tablefootmark{b}     & 0.36      & 0.33\tablefootmark{c} \\
    \citet{IdaEtal2016}\tablefootmark{d}            & 0.46      & 0.40 \\
    \hline
\end{tabular}
\tablefoot{In comparing the expressions we adopt the shear-dominated velocity regime in the 2D limit (massive planets). \\
    \tablefoottext{a}{Using their Equations (1), (29) and (31).}\\
    \tablefoottext{b}{In the 2D limit the expression followed \citet{LambrechtsJohansen2014}. \citet{BitschEtal2015} also adopts these expressions. } \\
    \tablefoottext{c}{Did not obtain the correct leading term in the case when the approach velocity was given by the Keplerian shear.} \\
    \tablefoottext{d}{Their Equation (59).}
}
\label{tab:collected}
\end{table}
\section{Discussion}
\label{sec:discussion}
\Tb{collected} compiles expressions for $\varepsilon$ that have been used in, or derived from, recent studies. For simplicity, we ignore the $f_\mathrm{set}$ term and list the numerical prefactor belonging to the leading term for the 3D and the 2D limits. In the 2D limit the shear-dominated velocity regime has been adopted (valid for large planets), while in the 3D limit the expression is independent of $\Delta v$. We emphasize that this work gives the correct numerical prefactor, as it is calibrated against numerical simulations.
On the other hand, the existing literature expressions often employed scaling arguments, where typically the 3D rate is estimated to be a fraction $b_\mathrm{set}/H_P$ of the 2D rate where $b_\mathrm{set}$ is the pebble accretion impact parameter. It is therefore quite remarkable that the existing literature prefactors lie so close to our calculated values. Nevertheless, in the 2D-limit the literature expression turn out to be too high, up to 50\%. This may still be significant, since a factor of two difference means that planet formation by pebble accretion takes twice as long and requires twice the number of pebbles.

Like our study \citet{Chambers2014} considered the cases where planet eccentricity (inclinations) dominate and accounts for the velocity effect of turbulence. However, he only considered the turbulent rms velocity for the relative motion between planet and pebble, which suppresses pebble accretion exponentially ($f_\mathrm{set}\gg1$) once $\sigma\gg v_\ast$. This is illustrated in \fg{fset} by the gray curve, where the settling fraction is plotted as function of planet mass. Parameters are chosen such that turbulent rms velocities $\sim$$\alpha_z^{1/2}h_\mathrm{gas} v_K$ are similar to the laminar headwind velocity, $\eta v_K$. When $f_\mathrm{set}$ is calculated by adding in quadrature the laminar and turbulent rms velocities (as in \eq{fsetI}) it ensures exponential behavior at low $q_p$.
However, accounting for a velocity distribution changes the functional behavior of $f_\mathrm{set}$ and, for the adopted parameters, we obtain the counter intuitive result that turbulence increases $f_\mathrm{set}$ (at low $q_p$; solid curve).
Accounting for the distribution, there are always a few particles slow enough for the settling mechanism to operate. This finding is important especially in the outer disk, where even at $f_\mathrm{set}\sim10^{-3}$ settling interactions dominate over ballistic interactions.

\begin{figure}[t]
    \centering
    \includegraphics[width=0.45\textwidth]{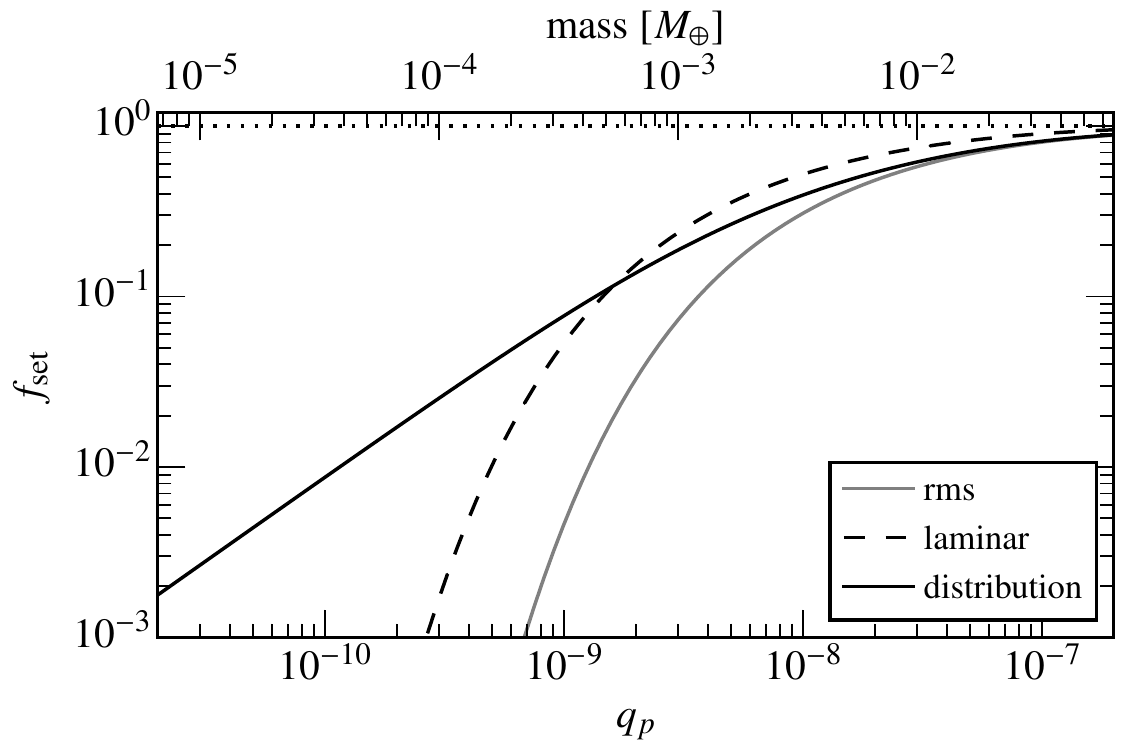}
    \caption{Settling fraction $f_\mathrm{set}$ as function of planet mass for $\alpha_z=10^{-2}$, $h_\mathrm{gas}=0.05$, $\eta=5\times10^{-3}$, $\tau_s=0.1$ and $M_\star=1\,M_\odot$. Turbulence is assumed to be isotropic and the correlation time $t_\mathrm{corr}=\Omega^{-1}$. Curves give $f_\mathrm{set}$ when it is calculated based on the laminar-only velocity (dashed), turbulent rms velocity (gray), and for a distribution of velocities (solid).}
\label{fig:fset}
\end{figure}
Recently, \citet{XuEtal2017} measured pebble accretion rates from laminar and MRI-turbulent hydrodynamical simulations. They expressed their result in terms of a dimensionless quantity $k_\mathrm{abs}$, which is the ratio of the mass accretion rate ($\dot{M}$) normalized to the Hill accretion rate $3R_H^2 \Omega \Sigma_P$. In these units, our expressions for the 2D and 3D efficiencies (\eq{eps-final}) read:\footnote{$k_\mathrm{abs}$ follows from \eq{eps-final} by multiplication by $2\pi r v_r/3R_H^2 \Omega$ where $v_r=2\eta v_K \tau_s$. In \eq{kabs-2D} we also substituted $v_\mathrm{cir}$ (\eq{vcir}) for $\Delta v$.}
\begin{subequations}
    \label{eq:kabs}
\begin{equation}
    \label{eq:kabs-2D}
    k_\mathrm{abs,2D}
    = 2.8 \tau_s^{1/2} \sqrt{\left( \left( \frac{q_p}{\eta^3} \right)^{1/3}  +a_\mathrm{cir} \tau_s \left(\frac{q_p}{\eta^3}\right)^{4/3} \right)^{-1} +a_\mathrm{sh} \tau_s^{1/3}}
\end{equation}
\begin{equation}
    \label{eq:kabs-3D}
    k_\mathrm{abs,3D}
    = 3.4 \frac{q_p^{1/3}\tau_s}{h_P}
    = 3.4 \left( \frac{q_p}{h_\mathrm{gas}^3} \right)^{1/3} \frac{\tau_s}{h_P/h_\mathrm{gas}}
\end{equation}
\end{subequations}
where we did insert $\Delta v$ in \eq{kabs-2D}, but have for clarity omitted the $f_\mathrm{set}$ modulation factor.

As a note in passing, \eq{kabs-2D} depends, apart from $\tau_s$, only on the quantity $q_p/\eta^3$. The reason is that the pebble equation of motion, expressed in Hill units, only contains this parameter\footnote{See \citet{OrmelKlahr2010}, where the parameter, denoted $\zeta_w$, is expressed as the ratio of the Hill velocity to the disk headwind, $\zeta_w= (3\eta^3/q_p)^{1/3}$.}. Likewise, the 3D-rates also depend on a single, but different, parameter (assuming $h_\mathrm{eff}$ can be quantified in terms of the stopping time). In the general case, then, two parameters (or three if $\tau_s$ is included) are necessary to calculate the Hill accretion rate. In their local shearing box simulations, \citet{XuEtal2017} choose to fix the thermal mass $q_p/h_\mathrm{gas}^3$ and the quantity $\eta/h_\mathrm{gas}$. For the Hill accretion rate ($k_\mathrm{abs}$), the problem is then fully specified. However, to calculate the efficiencies ($\varepsilon$) the degeneracy that existed among  $q_p$, $\eta$, and $h_\mathrm{gas}$ is broken -- $q_p$, $\eta$, and $h_\mathrm{gas}$ each need to be specified. This is because the accretion rate is a local quantity, whereas in order to calculate $\varepsilon$ the radial pebble flux must be known, \ie the disk circumference should be specified.

\begin{figure}[t]
    \includegraphics[width=0.45\textwidth]{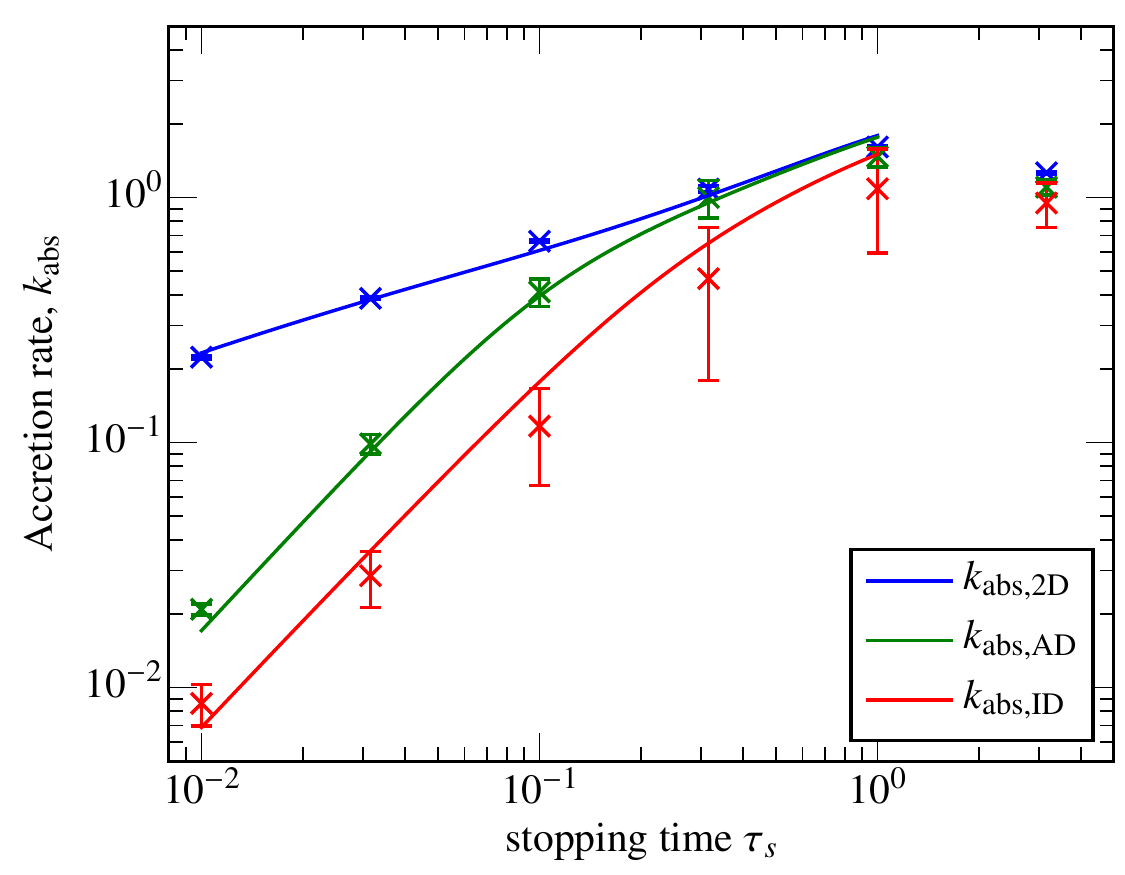}
    \caption{Hill-normalized accretion rates for a planet of thermal mass $q_p/h_\mathrm{gas}^3=3\times10^{-3}$ and $\eta/h_\mathrm{gas}=0.1$. Data points give $k_\mathrm{abs}$ obtained from hydrodynamical laminar (2D), ideal MRI (ID), and resistive (ambipolar diffusion) MRI (AD) simulations reported by \citet{XuEtal2017}. Solid lines give our fits, where we adopt $\sigma$ from the same simulations. Data kindly provided by Ziyan Xu.}
\label{fig:xu}
\end{figure}
In \fg{xu} symbols correspond to the runs conducted by \citet[][their Figure 4a]{XuEtal2017} for a thermal mass of $q_p/h_\mathrm{gas}^3=3\times10^{-3}$ and a headwind parameter of $\eta=0.1h_\mathrm{gas}$. Blue symbols correspond to the hydrodynamic runs (non-turbulent), green symbols to the ambipolar diffusion runs, and red symbols to the ideal MRI runs. As expected, the accretion rate decreases in the turbulent case and more so in the ideal MRI run than in the AD-restive run. Solid curves give the corresponding $k_\mathrm{abs}$ obtained from \eq{kabs}, $f_\mathrm{set}$, and the 2D and 3D averaging formula (\eq{eps-combine}).\footnote{In evaluating the expressions we used the turbulent diffusivities reported by \citet{XuEtal2017}: $\alpha_z = 7.8\times10^{-4}$ (ambipolar diffusion) and $\alpha_z=4.4\times10^{-3}$ (ideal MRI). Similarly, we use the (midplane) rms gas velocities from these simulations (Ziyan Xu, priv. comm.).}
Fits are plotted until $\tau_s=1$ beyond which they loose their validity.

The results of our analytical model agree well with \citet{XuEtal2017}. For the turbulent runs, we find that the reduction of the accretion rate is mostly due to turbulent diffusion (turbulence lofting pebbles away from the midplane) rather than the turbulent velocity effect.  This explains why the ideal MRI run features lower accretion rates than the ambipolar diffusion runs. Towards $\tau_s=0.1$--$1$ we see that the ambipolar diffusion run and the hydrodynamic run converge, as the accretion becomes 2D and turbulent diffusion does not enter $k_\mathrm{abs}$. However, we also find that the ideal MRI run does not entirely converge on the hydrodynamic run. This can be attributed to the turbulent velocity effect; $f_\mathrm{set} \lesssim 1$ because $\sigma \gtrsim v_\ast$ for the $\tau_s\simeq0.1$--1 particles in the MRI run. In other words, settling is marginally failing. We anticipate a much stronger reduction for either smaller planets or more vigorous turbulence.

\section{Summary}
\label{sec:summary}
In this work we have developed a general framework for the stochastic equation of motion (SEOM) of pebble-sized particles. The SEOM is described by the particle's stopping time, by the gas diffusivity $D_\mathrm{gas}$ and correlation time $t_\mathrm{corr}$, and by gravitational forces.  Using the SEOM we have investigated the vertical transport of particles in disks.
\begin{enumerate}
    \item From the SEOM we obtain the strong coupling approximation (SCA) by taking the limit of $t_\mathrm{stop}\rightarrow0$, as was used in previous studies \citep{Ciesla2010,ZsomEtal2011}. For small $t_\mathrm{stop}$ and $t_\mathrm{corr}$ the SEOM becomes consistent with the SCA.
    \item Exploring the effect of a vertically dependent diffusivity, $\alpha_z(z)$, such that the midplane diffusivity $\alpha_\mathrm{mid} \ll \tau_s$ and the surface diffusivity $\alpha_\mathrm{surface} \gg \tau_s$, we find a two component distribution for the pebbles. Although most of the pebbles are concentrated in the midplane (following the scaleheight given by $\alpha_\mathrm{mid}$), a small fraction of pebbles are distributed according to the gas scaleheight. Then, high-velocity collisions among these pebbles could explain the prolonged presence of small, micron-sized grains in the disk surface.
\end{enumerate}

As its main application, we have integrated pebble trajectories to find the pebble accretion efficiency $\varepsilon$.
\begin{enumerate}[resume*]
    \item Compared to the laminar disk, turbulence reduces $\varepsilon$ in two ways: it diminishes the local density of pebbles at the midplane through diffusion and it increases the rms velocity of particles. The latter becomes important for low planet masses, where fulfilling the settling condition becomes more difficult.
    \item Because of the disk geometry, pebble accretion is more efficient in the inner disk. From an efficiency perspective, accretion around low-mass stars is also favored because pebble capture radii are larger around low-mass stars.
    \item Together with the 2D expressions already derived in Paper I, we have formulated a general prescription for $\varepsilon$ as a function of pebble properties (aerodynamical size), planet properties (mass, inclination, eccentricity), and disk properties (pressure profile, gas density, turbulence properties). This prescription is summarized in \se{fit}.
\end{enumerate}

Finally, we remark that $\varepsilon$ has been defined with respect to a single planet. A small $\varepsilon$ therefore does not necessarily imply that pebble accretion is globally inefficient. Indeed, in the case of a planetesimal belt (or multiple planets) the total filtering efficiency may well reach unity, whereas the individual $\varepsilon \ll1$ \citep{GuillotEtal2014}. This raises the question of how pebble accretion proceeds when multiple seeds are present.

While it is likely to result in a more efficient pebble sweep-up, a multi-seed scenario could also suppress planet growth because of the mutual dynamical excitation among the protoplanets \citep{LevisonEtal2015}. In a following work, we will investigate the efficacy of pebble accretion when pebbles interact with a narrow planetesimal belt (Liu et al., in prep). In such cases, the combined planetesimal and pebble coagulation can best be studied with N-body techniques, where the pebble accretion rate on the N bodies is obtained using the prescription for $\varepsilon$ summarized in \se{fit}.

\begin{acknowledgements}
The authors are supported by the Netherlands Organization for Scientific Research (NWO; VIDI project 639.042.422).  We thank Xuening Bai and Ziyan Xu for sharing and discussing their simulations, and Ramon Brasser and Sebastiaan Krijt for proofreading the manuscript. The comments from the referee improved the clarity of the manuscript.
\end{acknowledgements}

\bibliographystyle{aa}
\bibliography{ads,arXiv}

\begin{thebibliography}{91}
\expandafter\ifx\csname natexlab\endcsname\relax\def\natexlab#1{#1}\fi

\bibitem[{{Ansdell} {et~al.}(2017){Ansdell}, {Williams}, {Manara}, {Miotello},
  {Facchini}, {van der Marel}, {Testi}, \& {van Dishoeck}}]{AnsdellEtal2017}
{Ansdell}, M., {Williams}, J.~P., {Manara}, C.~F., {et~al.} 2017, \aj, 153, 240

\bibitem[{{Bai} \& {Stone}(2011)}]{BaiStone2011}
{Bai}, X.-N. \& {Stone}, J.~M. 2011, \apj, 736, 144

\bibitem[{{Bai} {et~al.}(2016){Bai}, {Ye}, {Goodman}, \& {Yuan}}]{BaiEtal2016}
{Bai}, X.-N., {Ye}, J., {Goodman}, J., \& {Yuan}, F. 2016, \apj, 818, 152

\bibitem[{{Balbus} \& {Hawley}(1991)}]{BalbusHawley1991}
{Balbus}, S.~A. \& {Hawley}, J.~F. 1991, \apj, 376, 214

\bibitem[{{Banzatti} {et~al.}(2015){Banzatti}, {Pinilla}, {Ricci},
  {Pontoppidan}, {Birnstiel}, \& {Ciesla}}]{BanzattiEtal2015}
{Banzatti}, A., {Pinilla}, P., {Ricci}, L., {et~al.} 2015, \apjl, 815, L15

\bibitem[{{Birnstiel} {et~al.}(2010){Birnstiel}, {Dullemond}, \&
  {Brauer}}]{BirnstielEtal2010i}
{Birnstiel}, T., {Dullemond}, C.~P., \& {Brauer}, F. 2010, \aap, 513, A79

\bibitem[{{Birnstiel} {et~al.}(2011){Birnstiel}, {Ormel}, \&
  {Dullemond}}]{BirnstielEtal2011}
{Birnstiel}, T., {Ormel}, C.~W., \& {Dullemond}, C.~P. 2011, \aap, 525, A11

\bibitem[{{Bitsch} {et~al.}(2015){Bitsch}, {Lambrechts}, \&
  {Johansen}}]{BitschEtal2015}
{Bitsch}, B., {Lambrechts}, M., \& {Johansen}, A. 2015, \aap, 582, A112

\bibitem[{{Bitsch} {et~al.}(2018{\natexlab{a}}){Bitsch}, {Lambrechts}, \&
  {Johansen}}]{BitschEtal2018}
{Bitsch}, B., {Lambrechts}, M., \& {Johansen}, A. 2018{\natexlab{a}}, \aap,
  609, C2

\bibitem[{{Bitsch} {et~al.}(2018{\natexlab{b}}){Bitsch}, {Morbidelli},
  {Johansen}, {Lega}, {Lambrechts}, \& {Crida}}]{BitschEtal2018i}
{Bitsch}, B., {Morbidelli}, A., {Johansen}, A., {et~al.} 2018{\natexlab{b}},
  ArXiv e-prints:1801.02341

\bibitem[{{Brouwers} {et~al.}(2018){Brouwers}, {Vazan}, \&
  {Ormel}}]{BrouwersEtal2018}
{Brouwers}, M.~G., {Vazan}, A., \& {Ormel}, C.~W. 2018, \aap, 611, A65

\bibitem[{{Carballido} {et~al.}(2011){Carballido}, {Bai}, \&
  {Cuzzi}}]{CarballidoEtal2011}
{Carballido}, A., {Bai}, X.-N., \& {Cuzzi}, J.~N. 2011, \mnras, 415, 93

\bibitem[{{Chambers}(2017)}]{Chambers2017}
{Chambers}, J. 2017, \apj, 849, 30

\bibitem[{{Chambers}(2014)}]{Chambers2014}
{Chambers}, J.~E. 2014, \icarus, 233, 83

\bibitem[{{Charnoz} {et~al.}(2011){Charnoz}, {Fouchet}, {Aleon}, \&
  {Moreira}}]{CharnozEtal2011}
{Charnoz}, S., {Fouchet}, L., {Aleon}, J., \& {Moreira}, M. 2011, \apj, 737, 33

\bibitem[{{Chatterjee} \& {Tan}(2014)}]{ChatterjeeTan2014}
{Chatterjee}, S. \& {Tan}, J.~C. 2014, \apj, 780, 53

\bibitem[{{Ciesla}(2010)}]{Ciesla2010}
{Ciesla}, F.~J. 2010, \apj, 723, 514

\bibitem[{{Cuzzi} {et~al.}(2001){Cuzzi}, {Hogan}, {Paque}, \&
  {Dobrovolskis}}]{CuzziEtal2001}
{Cuzzi}, J.~N., {Hogan}, R.~C., {Paque}, J.~M., \& {Dobrovolskis}, A.~R. 2001,
  \apj, 546, 496

\bibitem[{{Cuzzi} \& {Zahnle}(2004)}]{CuzziZahnle2004}
{Cuzzi}, J.~N. \& {Zahnle}, K.~J. 2004, \apj, 614, 490

\bibitem[{{Dr{\c a}{\.z}kowska} \& {Alibert}(2017)}]{DrazkowskaAlibert2017}
{Dr{\c a}{\.z}kowska}, J. \& {Alibert}, Y. 2017, \aap, 608, A92

\bibitem[{{Dubrulle} {et~al.}(1995){Dubrulle}, {Morfill}, \&
  {Sterzik}}]{DubrulleEtal1995}
{Dubrulle}, B., {Morfill}, G., \& {Sterzik}, M. 1995, Icarus, 114, 237

\bibitem[{{Dullemond} \& {Dominik}(2005)}]{DullemondDominik2005}
{Dullemond}, C.~P. \& {Dominik}, C. 2005, \aap, 434, 971

\bibitem[{{Flaherty} {et~al.}(2017){Flaherty}, {Hughes}, {Rose}, {Simon}, {Qi},
  {Andrews}, {K{\'o}sp{\'a}l}, {Wilner}, {Chiang}, {Armitage}, \&
  {Bai}}]{FlahertyEtal2017}
{Flaherty}, K.~M., {Hughes}, A.~M., {Rose}, S.~C., {et~al.} 2017, \apj, 843,
  150

\bibitem[{{Flaherty} {et~al.}(2015){Flaherty}, {Hughes}, {Rosenfeld},
  {Andrews}, {Chiang}, {Simon}, {Kerzner}, \& {Wilner}}]{FlahertyEtal2015}
{Flaherty}, K.~M., {Hughes}, A.~M., {Rosenfeld}, K.~A., {et~al.} 2015, \apj,
  813, 99

\bibitem[{{Flock} {et~al.}(2017){Flock}, {Fromang}, {Turner}, \&
  {Benisty}}]{FlockEtal2017}
{Flock}, M., {Fromang}, S., {Turner}, N.~J., \& {Benisty}, M. 2017, \apj, 835,
  230

\bibitem[{{Gammie}(1996)}]{Gammie1996}
{Gammie}, C.~F. 1996, \apj, 457, 355

\bibitem[{{Gressel}(2017)}]{Gressel2017}
{Gressel}, O. 2017, in Journal of Physics Conference Series, Vol. 837, Journal
  of Physics Conference Series, 012008

\bibitem[{{Gressel} {et~al.}(2011){Gressel}, {Nelson}, \&
  {Turner}}]{GresselEtal2011}
{Gressel}, O., {Nelson}, R.~P., \& {Turner}, N.~J. 2011, \mnras, 415, 3291

\bibitem[{{Gressel} {et~al.}(2012){Gressel}, {Nelson}, \&
  {Turner}}]{GresselEtal2012}
{Gressel}, O., {Nelson}, R.~P., \& {Turner}, N.~J. 2012, \mnras, 422, 1140

\bibitem[{{Guillot} {et~al.}(2014){Guillot}, {Ida}, \&
  {Ormel}}]{GuillotEtal2014}
{Guillot}, T., {Ida}, S., \& {Ormel}, C.~W. 2014, \aap, 572, A72

\bibitem[{{Guilloteau} {et~al.}(2012){Guilloteau}, {Dutrey}, {Wakelam},
  {Hersant}, {Semenov}, {Chapillon}, {Henning}, \&
  {Pi{\'e}tu}}]{GuilloteauEtal2012}
{Guilloteau}, S., {Dutrey}, A., {Wakelam}, V., {et~al.} 2012, \aap, 548, A70

\bibitem[{{Homann} {et~al.}(2016){Homann}, {Guillot}, {Bec}, {Ormel}, {Ida}, \&
  {Tanga}}]{HomannEtal2016}
{Homann}, H., {Guillot}, T., {Bec}, J., {et~al.} 2016, \aap, 589, A129

\bibitem[{{Hottovy} {et~al.}(2015){Hottovy}, {McDaniel}, {Volpe}, \&
  {Wehr}}]{HottovyEtal2015}
{Hottovy}, S., {McDaniel}, A., {Volpe}, G., \& {Wehr}, J. 2015, Communications
  in Mathematical Physics, 336, 1259

\bibitem[{{Hughes} {et~al.}(2011){Hughes}, {Wilner}, {Andrews}, {Qi}, \&
  {Hogerheijde}}]{HughesEtal2011}
{Hughes}, A.~M., {Wilner}, D.~J., {Andrews}, S.~M., {Qi}, C., \& {Hogerheijde},
  M.~R. 2011, \apj, 727, 85

\bibitem[{{Ida} \& {Guillot}(2016)}]{IdaGuillot2016}
{Ida}, S. \& {Guillot}, T. 2016, \aap, 596, L3

\bibitem[{{Ida} {et~al.}(2008){Ida}, {Guillot}, \& {Morbidelli}}]{IdaEtal2008}
{Ida}, S., {Guillot}, T., \& {Morbidelli}, A. 2008, \apj, 686, 1292

\bibitem[{{Ida} {et~al.}(2016){Ida}, {Guillot}, \& {Morbidelli}}]{IdaEtal2016}
{Ida}, S., {Guillot}, T., \& {Morbidelli}, A. 2016, \aap, 591, A72

\bibitem[{{Johansen} {et~al.}(2006){Johansen}, {Klahr}, \&
  {Mee}}]{JohansenEtal2006i}
{Johansen}, A., {Klahr}, H., \& {Mee}, A.~J. 2006, \mnras, 370, L71

\bibitem[{{Johansen} \& {Lambrechts}(2017)}]{JohansenLambrechts2017}
{Johansen}, A. \& {Lambrechts}, M. 2017, Annual Review of Earth and Planetary
  Sciences, 45, 359

\bibitem[{{Johansen} {et~al.}(2015){Johansen}, {Mac Low}, {Lacerda}, \&
  {Bizzarro}}]{JohansenEtal2015}
{Johansen}, A., {Mac Low}, M.-M., {Lacerda}, P., \& {Bizzarro}, M. 2015,
  Science Advances, 1, 15109

\bibitem[{{Juh{\'a}sz} {et~al.}(2010){Juh{\'a}sz}, {Bouwman}, {Henning},
  {Acke}, {van den Ancker}, {Meeus}, {Dominik}, {Min}, {Tielens}, \&
  {Waters}}]{JuhaszEtal2010}
{Juh{\'a}sz}, A., {Bouwman}, J., {Henning}, T., {et~al.} 2010, \apj, 721, 431

\bibitem[{{Kanagawa} {et~al.}(2017){Kanagawa}, {Ueda}, {Muto}, \&
  {Okuzumi}}]{KanagawaEtal2017}
{Kanagawa}, K.~D., {Ueda}, T., {Muto}, T., \& {Okuzumi}, S. 2017, \apj, 844,
  142

\bibitem[{{Kobayashi} {et~al.}(2010){Kobayashi}, {Tanaka}, {Krivov}, \&
  {Inaba}}]{KobayashiEtal2010}
{Kobayashi}, H., {Tanaka}, H., {Krivov}, A.~V., \& {Inaba}, S. 2010, \icarus,
  209, 836

\bibitem[{{Kobayashi} {et~al.}(2016){Kobayashi}, {Tanaka}, \&
  {Okuzumi}}]{KobayashiEtal2016}
{Kobayashi}, H., {Tanaka}, H., \& {Okuzumi}, S. 2016, \apj, 817, 105

\bibitem[{{Krijt} \& {Ciesla}(2016)}]{KrijtCiesla2016}
{Krijt}, S. \& {Ciesla}, F.~J. 2016, \apj, 822, 111

\bibitem[{{Lambrechts} \& {Johansen}(2012)}]{LambrechtsJohansen2012}
{Lambrechts}, M. \& {Johansen}, A. 2012, \aap, 544, A32

\bibitem[{{Lambrechts} \& {Johansen}(2014)}]{LambrechtsJohansen2014}
{Lambrechts}, M. \& {Johansen}, A. 2014, \aap, 572, A107

\bibitem[{{Lambrechts} {et~al.}(2014){Lambrechts}, {Johansen}, \&
  {Morbidelli}}]{LambrechtsEtal2014}
{Lambrechts}, M., {Johansen}, A., \& {Morbidelli}, A. 2014, \aap, 572, A35

\bibitem[{{Levison} {et~al.}(2015){Levison}, {Kretke}, \&
  {Duncan}}]{LevisonEtal2015}
{Levison}, H.~F., {Kretke}, K.~A., \& {Duncan}, M.~J. 2015, \nat, 524, 322

\bibitem[{{Liu} \& {Ormel}(2018)}]{LiuOrmel2018}
{Liu}, B. \& {Ormel}, C.~W. 2018, ArXiv e-prints:1803.06149 (Paper I)

\bibitem[{{Miotello} {et~al.}(2017){Miotello}, {van Dishoeck}, {Williams},
  {Ansdell}, {Guidi}, {Hogerheijde}, {Manara}, {Tazzari}, {Testi}, {van der
  Marel}, \& {van Terwisga}}]{MiotelloEtal2017}
{Miotello}, A., {van Dishoeck}, E.~F., {Williams}, J.~P., {et~al.} 2017, \aap,
  599, A113

\bibitem[{{Morbidelli} {et~al.}(2015){Morbidelli}, {Lambrechts}, {Jacobson}, \&
  {Bitsch}}]{MorbidelliEtal2015}
{Morbidelli}, A., {Lambrechts}, M., {Jacobson}, S., \& {Bitsch}, B. 2015,
  \icarus, 258, 418

\bibitem[{{Nakagawa} {et~al.}(1986){Nakagawa}, {Sekiya}, \&
  {Hayashi}}]{NakagawaEtal1986}
{Nakagawa}, Y., {Sekiya}, M., \& {Hayashi}, C. 1986, Icarus, 67, 375

\bibitem[{{Nelson} \& {Gressel}(2010)}]{NelsonGressel2010}
{Nelson}, R.~P. \& {Gressel}, O. 2010, \mnras, 409, 639

\bibitem[{{Nelson} {et~al.}(2013){Nelson}, {Gressel}, \&
  {Umurhan}}]{NelsonEtal2013}
{Nelson}, R.~P., {Gressel}, O., \& {Umurhan}, O.~M. 2013, \mnras, 435, 2610

\bibitem[{{Okuzumi} \& {Ormel}(2013)}]{OkuzumiOrmel2013}
{Okuzumi}, S. \& {Ormel}, C.~W. 2013, \apj, 771, 43

\bibitem[{{Ormel}(2013)}]{Ormel2013}
{Ormel}, C.~W. 2013, \mnras, 428, 3526

\bibitem[{{Ormel}(2017)}]{Ormel2017}
{Ormel}, C.~W. 2017, in Astrophysics and Space Science Library, Vol. 445,
  Astrophysics and Space Science Library, ed. M.~{Pessah} \& O.~{Gressel}, 197

\bibitem[{{Ormel} \& {Cuzzi}(2007)}]{OrmelCuzzi2007}
{Ormel}, C.~W. \& {Cuzzi}, J.~N. 2007, \aap, 466, 413

\bibitem[{{Ormel} \& {Klahr}(2010)}]{OrmelKlahr2010}
{Ormel}, C.~W. \& {Klahr}, H.~H. 2010, \aap, 520, A43

\bibitem[{{Ormel} {et~al.}(2017){Ormel}, {Liu}, \&
  {Schoonenberg}}]{OrmelEtal2017}
{Ormel}, C.~W., {Liu}, B., \& {Schoonenberg}, D. 2017, \aap, 604, A1

\bibitem[{{Ormel} \& {Okuzumi}(2013)}]{OrmelOkuzumi2013}
{Ormel}, C.~W. \& {Okuzumi}, S. 2013, \apj, 771, 44

\bibitem[{{Paardekooper} {et~al.}(2013){Paardekooper}, {Rein}, \&
  {Kley}}]{PaardekooperEtal2013}
{Paardekooper}, S.-J., {Rein}, H., \& {Kley}, W. 2013, \mnras, 434, 3018

\bibitem[{{Pan} \& {Padoan}(2010)}]{PanPadoan2010}
{Pan}, L. \& {Padoan}, P. 2010, Journal of Fluid Mechanics, 661, 73

\bibitem[{{Picogna} {et~al.}(2018){Picogna}, {Stoll}, \&
  {Kley}}]{PicognaEtal2018}
{Picogna}, G., {Stoll}, M.~H.~R., \& {Kley}, W. 2018, ArXiv e-prints:1803.08730

\bibitem[{{Pinilla} {et~al.}(2012){Pinilla}, {Birnstiel}, {Ricci}, {Dullemond},
  {Uribe}, {Testi}, \& {Natta}}]{PinillaEtal2012}
{Pinilla}, P., {Birnstiel}, T., {Ricci}, L., {et~al.} 2012, \aap, 538, A114

\bibitem[{{Pinte} {et~al.}(2016){Pinte}, {Dent}, {M{\'e}nard}, {Hales}, {Hill},
  {Cortes}, \& {de Gregorio-Monsalvo}}]{PinteEtal2016}
{Pinte}, C., {Dent}, W.~R.~F., {M{\'e}nard}, F., {et~al.} 2016, \apj, 816, 25

\bibitem[{{Pollack} {et~al.}(1996){Pollack}, {Hubickyj}, {Bodenheimer},
  {Lissauer}, {Podolak}, \& {Greenzweig}}]{PollackEtal1996}
{Pollack}, J.~B., {Hubickyj}, O., {Bodenheimer}, P., {et~al.} 1996, Icarus,
  124, 62

\bibitem[{{Rein} \& {Papaloizou}(2009)}]{ReinPapaloizou2009}
{Rein}, H. \& {Papaloizou}, J.~C.~B. 2009, \aap, 497, 595

\bibitem[{{Ros} \& {Johansen}(2013)}]{RosJohansen2013}
{Ros}, K. \& {Johansen}, A. 2013, \aap, 552, A137

\bibitem[{{Safronov}(1969)}]{Safronov1969}
{Safronov}, V.~S. 1969, {Evolution of the Protoplanetary Cloud and Formation of
  Earth and the Planets}, ed. V.~S. Safronov (Moscow: Nauka. Transl. 1972 NASA
  Tech. F-677)

\bibitem[{{Sano} {et~al.}(2004){Sano}, {Inutsuka}, {Turner}, \&
  {Stone}}]{SanoEtal2004}
{Sano}, T., {Inutsuka}, S.-i., {Turner}, N.~J., \& {Stone}, J.~M. 2004, \apj,
  605, 321

\bibitem[{{Schoonenberg} \& {Ormel}(2017)}]{SchoonenbergOrmel2017}
{Schoonenberg}, D. \& {Ormel}, C.~W. 2017, \aap, 602, A21

\bibitem[{{Shakura} \& {Sunyaev}(1973)}]{ShakuraSunyaev1973}
{Shakura}, N.~I. \& {Sunyaev}, R.~A. 1973, \aap, 24, 337

\bibitem[{{Smoluchowski}(1916)}]{Smoluchowski1916}
{Smoluchowski}, M.~V. 1916, Zeitschrift f\"ur Physik, 17, 557

\bibitem[{{Stoll} \& {Kley}(2014)}]{StollKley2014}
{Stoll}, M.~H.~R. \& {Kley}, W. 2014, \aap, 572, A77

\bibitem[{{Stoll} {et~al.}(2017){Stoll}, {Kley}, \& {Picogna}}]{StollEtal2017}
{Stoll}, M.~H.~R., {Kley}, W., \& {Picogna}, G. 2017, \aap, 599, L6

\bibitem[{{Suzuki} {et~al.}(2016){Suzuki}, {Ogihara}, {Morbidelli}, {Crida}, \&
  {Guillot}}]{SuzukiEtal2016}
{Suzuki}, T.~K., {Ogihara}, M., {Morbidelli}, A., {Crida}, A., \& {Guillot}, T.
  2016, \aap, 596, A74

\bibitem[{{Takeuchi} \& {Lin}(2002)}]{TakeuchiLin2002}
{Takeuchi}, T. \& {Lin}, D.~N.~C. 2002, \apj, 581, 1344

\bibitem[{{Tanaka} {et~al.}(2005){Tanaka}, {Himeno}, \& {Ida}}]{TanakaEtal2005}
{Tanaka}, H., {Himeno}, Y., \& {Ida}, S. 2005, \apj, 625, 414

\bibitem[{{Teague} {et~al.}(2016){Teague}, {Guilloteau}, {Semenov}, {Henning},
  {Dutrey}, {Pi{\'e}tu}, {Birnstiel}, {Chapillon}, {Hollenbach}, \&
  {Gorti}}]{TeagueEtal2016}
{Teague}, R., {Guilloteau}, S., {Semenov}, D., {et~al.} 2016, \aap, 592, A49

\bibitem[{{Uhlenbeck} \& {Ornstein}(1930)}]{UhlenbeckOrnstein1930}
{Uhlenbeck}, G.~E. \& {Ornstein}, L.~S. 1930, Physical Review, 36, 823

\bibitem[{{van Kampen}(1992)}]{vanKampen1992}
{van Kampen}, N.~G. 1992, {Stochastic Processes in Physics and Chemistry}

\bibitem[{{Visser} \& {Ormel}(2016)}]{VisserOrmel2016}
{Visser}, R.~G. \& {Ormel}, C.~W. 2016, \aap, 586, A66

\bibitem[{{V\"olk} {et~al.}(1980){V\"olk}, {Jones}, {Morfill}, \&
  {Roeser}}]{VoelkEtal1980}
{V\"olk}, H.~J., {Jones}, F.~C., {Morfill}, G.~E., \& {Roeser}, S. 1980, \aap,
  85, 316

\bibitem[{{Weidenschilling}(1977)}]{Weidenschilling1977}
{Weidenschilling}, S.~J. 1977, \mnras, 180, 57

\bibitem[{{Weidenschilling} \& {Davis}(1985)}]{WeidenschillingDavis1985}
{Weidenschilling}, S.~J. \& {Davis}, D.~R. 1985, Icarus, 62, 16

\bibitem[{{Xu} {et~al.}(2017){Xu}, {Bai}, \& {Murray-Clay}}]{XuEtal2017}
{Xu}, Z., {Bai}, X.-N., \& {Murray-Clay}, R.~A. 2017, \apj, 847, 52

\bibitem[{{Youdin} \& {Lithwick}(2007)}]{YoudinLithwick2007}
{Youdin}, A.~N. \& {Lithwick}, Y. 2007, Icarus, 192, 588

\bibitem[{{Zhu} {et~al.}(2015){Zhu}, {Stone}, \& {Bai}}]{ZhuEtal2015}
{Zhu}, Z., {Stone}, J.~M., \& {Bai}, X.-N. 2015, \apj, 801, 81

\bibitem[{{Zsom} {et~al.}(2011){Zsom}, {Ormel}, {Dullemond}, \&
  {Henning}}]{ZsomEtal2011}
{Zsom}, A., {Ormel}, C.~W., {Dullemond}, C.~P., \& {Henning}, T. 2011, \aap,
  534, A73

\end{thebibliography}

\appendix
\section{Proof of \Eq{xt}}
\label{app:Hottovy}
In deriving \eq{xt}, we follow the proof outlined by \citet{HottovyEtal2015}. The first step is to write \eq{eom-all} as:
\begin{subequations}
    \label{eq:Hottovy1}
    \begin{equation}
        \dd \bm{x} = \bm{v} \dd t
    \end{equation}
    \begin{equation}
        t_\mathrm{stop} \dd \bm{v} = \bm{F} \dd t -\boldsymbol{\upgamma} \bm{v} \dd t +\bm{\sigma} \dd W_t
    \end{equation}
\end{subequations}
In this equation, the stochastic variable $\zeta_t$ has been combined with the velocity into one vector $\bm{v} = (v, \zeta_t)$. Similarly, $\bm{x}$ contains an additional parameter, say $\upsilon$, but this is entirely dummy. Comparing with \eq{eom-all} we therefore have:
\begin{subequations}
\begin{equation}
    \bm{\tilde{F}} = \left( \begin{array}{cc}
    \bm{v}_\mathrm{gas} +\bm{v}_\mathrm{hs} +\bm{F}_g t_\mathrm{stop} \\
    0
\end{array}
    \right)
\end{equation}
\begin{equation}
    \bm{\upgamma} = \left( \begin{array}{cc}
        1   &   -\sqrt{D_z/t_\mathrm{corr}} \\
        0   &   t_\mathrm{stop}/t_\mathrm{corr} \\
\end{array}
    \right)
\end{equation}
\begin{equation}
    \bm{\sigma} = \left( \begin{array}{cc}
    0 \\
    \sqrt{2t_\mathrm{stop}^2/t_\mathrm{corr}}
\end{array}
    \right)
\end{equation}
\end{subequations}
It must be emphasized that we only treat here the $z$ coordinate. When the full 3D equation of motion is considered, $\bm{v}$ will be vector of length six and $\boldsymbol{\upgamma}$ a matrix of 36 elements, containing all entries of the diffusion tensor $D_{ij}$.

In the limit of $t_\mathrm{stop}\rightarrow0$, \citet{HottovyEtal2015} proves that $\bm{x}$ can be described by the SDE
\begin{equation}
    \label{eq:Hottovy-formal}
    \bm{x} = \left[ \boldsymbol{\upgamma}^{-1} \bm{\tilde{F}} +\bm{S} \right] \dd t +\boldsymbol{\upgamma}^{-1}\bm{\sigma} d\bm{W}_t
\end{equation}
where $\boldsymbol{\upgamma}^{-1}$ is the inverse of $\boldsymbol{\upgamma}$ and $\bm{S}$ is the noise-induced drift term -- a vector whose $i^\mathrm{th}$ component is defined
\begin{equation}
    \bm{S}_i = \sum_{j,l} \frac{\partial}{\partial x_l} \left[ (\boldsymbol{\upgamma}^{-1})_{ij} \right] \mathbf{J}_{jl}
\end{equation}
with $\mathbf{J}$ the solution of the Lyapunov equation
\begin{equation}
    \boldsymbol{\upgamma}\mathbf{J} +\mathbf{J}\boldsymbol{\upgamma}^\dagger = \bm{\sigma}\bm{\sigma}^\dagger.
\end{equation}
Since we consider here only the $z$ coordinate, \eq{Hottovy-formal} reduces too
\begin{equation}
    S_1 =
    \sum_j \frac{\partial}{\partial z} (\bm{\gamma}^{-1})_{1j} \mathbf{J}_\mathrm{j1}.
\end{equation}

The equations can now be solved. Inverting $\boldsymbol{\upgamma}$ we find
\begin{equation}
    \boldsymbol{\upgamma}^{-1} = \left( \begin{array}{cc}
        1   &   \sqrt{D_{zz}t_c}/t_\mathrm{stop} \\
        0   &   t_\mathrm{corr}/t_\mathrm{stop}
    \end{array} \right)
\end{equation}
Furthermore, solving the Lyapunov equation, we find
\begin{equation}
    \mathbf{J} = \left( \begin{array}{cc} \displaystyle
        \frac{t_\mathrm{stop}}{t_\mathrm{stop} +t_\mathrm{corr}} D(z)  & \displaystyle \frac{t_\mathrm{stop}}{t_\mathrm{stop}+t_\mathrm{corr}} \sqrt{D(z) t_\mathrm{corr}} \\ \displaystyle
        \frac{t_\mathrm{stop}}{t_\mathrm{stop}+t_\mathrm{corr}} \sqrt{D(z) t_\mathrm{corr}} & \displaystyle t_\mathrm{stop}
    \end{array} \right)
\end{equation}
with which the noise-induced drift term becomes
\begin{equation}
    \label{eq:S1x-diff}
    S_1(z)
    = \frac{1}{2} \frac{D'(z)}{1 +t_\mathrm{stop}/t_\mathrm{corr}}
\end{equation}
and we retrieve \eq{xt}.

\subsection{Stratonovich interpretation}
\label{sec:Stratonovich}
A feature peculiar to stochastic integrals is that equations of the form
\begin{equation}
    \label{eq:stoch-int-gen}
    \int_{t}^{t +\Delta t} B(x) \dd W_t
\end{equation}
are ill-defined when $B$ is a function of position. In contrast to ODEs, it does matter whether the integrand is evaluated at $t$ (\ie\ $B(x)=B(x[t])$ -- Ito's choice), at $t+\Delta t$, or whether we let
\begin{equation}
    \int_{t}^{\Delta t} B(x) \dd W_t \approx \frac{1}{2} B\left( \frac{x[t] +x[t+\Delta t]}{2} \right) \left( W_{t+\Delta t} -W_t \right)
\end{equation}
(Stratonovich' choice). In contrast to ODEs, these definitions will produce different results when $B$ is not constant \citep{vanKampen1992}. Specifically, with Stratonovich', rather than Ito's interpretation for the stochastic integral a term
\begin{equation}
    -\frac{1}{2} \frac{\partial}{\partial x} B \frac{\partial B}{\partial x}.
\end{equation}
should be added to the RHS of the Fokker-Planck \Eq{FP-Ito-0}.

In deriving \eq{xt}, as well as the conversion from \eq{xt} to \eq{FP-Ito} we have followed Ito's interpretation. On the other hand, under Stratonovich interpretation, the RHS of \eq{xt} and the RHS of \eq{FP-Ito} gain an additional term $-\frac{1}{2}D'$:
\begin{equation}
    \label{eq:xt-strat}
    \dd\bm{x}_t = \bm{F}t_\mathrm{stop} +\bm{v}_\mathrm{gas} +\bm{v}_\mathrm{hs}
    -\frac{t_\mathrm{stop} D_\mathrm{gas}'}{2(t_\mathrm{stop}+t_\mathrm{corr})} +\sqrt{2D} \dd W_t.
\end{equation}
Hence, the SDE for the strong coupling approximation depends on the interpretation rule (a fact not highlighted by \citealt{Ciesla2010} or \citealt{ZsomEtal2011}). Which form should we choose?

The underlying reason between the Ito and Stratonovich interpretations reflects the nature of the stochastic forcing \citep{vanKampen1992}. Ito's interpretation would hold, for example, when the stochastic forcing amounts to a series of infinitely short ``pulses'' with every pulse completely independent.  This interpretation is often used in finance However, it may be argued that for physical problems, where the correlation time is never really infinitely small, Stratonovich amounts to the more correct model \citep{vanKampen1992}.

More practically, these differences are expressed in our choice of the numerical integration scheme. When Ito's interpretation is adopted, the stochastic equation must be integrated with a corresponding numerical scheme, of which the Euler method is the simplest example. Similarly, the Stratonovich equation should be integrated with an appropriate numerical scheme, where the midpoint scheme (Heun's method) is the simplest example.  In our code, where we use an Runge-Kutta method, which is a generalization of a midpoint scheme, we therefore adopt \eq{xt-strat}, instead of \eq{xt3}.

\end{CJK*}
\end{document}